\documentclass[12pt]{article}
\usepackage{amsmath,amssymb}
\usepackage{graphicx,color}
\usepackage[linkcolor=blue,colorlinks=true]{hyperref}
\usepackage{cite,epsf}
\usepackage[bf]{caption}
\usepackage{subfig}
\usepackage[nottoc,notlot,notlof]{tocbibind}
\usepackage[title]{appendix}

\usepackage{tikz}
\usetikzlibrary{positioning}
\usetikzlibrary{intersections} 
\newlength{\PicScale}
\usepackage[active]{srcltx}
\usepackage{latexsym}
\usepackage{amsmath}
\usepackage{amsfonts}
\usepackage{mathrsfs}
\usepackage{dsfont}
\usepackage{amscd}
\usepackage{verbatim}
\usepackage{xcolor}
 \csname
@addtoreset\endcsname{equation}{section}

\usepackage{lipsum}       
\usepackage{xargs}                   
\usepackage[normalem]{ulem}

\usepackage[colorinlistoftodos,prependcaption,textsize=tiny]{todonotes}
\newcommandx{\unsure}[2][1=]{\todo[linecolor=red,backgroundcolor=red!25,bordercolor=red,#1]{#2}}
\newcommandx{\change}[2][1=]{\todo[linecolor=blue,backgroundcolor=blue!25,bordercolor=blue,#1]{#2}}
\newcommandx{\Sinfo}[1]{\todo[backgroundcolor=red!25,bordercolor=red,noline]{S:#1}}
\newcommandx{\Winfo}[1]{\todo[backgroundcolor=blue!25,bordercolor=blue,noline]{W:#1}}
\def\p{\partial}

\def\a{\alpha}
\def\b{\beta}
\def\g{\gamma}

\def\m{\mu}

\def\s{\sigma}

\def\u{\upsilon}

\def\qr{q\!\cdot\!r}

\newcommand{\be}{\begin{equation}}
\newcommand{\ee}{\end{equation}}
\newcommand{\ba}{\begin{aligned}}
\newcommand{\ea}{\end{aligned}}
\newcommand{\ben}{\begin{displaymath}}
\newcommand{\een}{\end{displaymath}}
\newcommand{\bea}{\begin{eqnarray}}
\newcommand{\eea}{\end{eqnarray}}

\def\12{\frac{1}{2}}

\def\eNm1{\overset{\scriptscriptstyle{(N-1)}}{e}}

\usepackage{xargs}

\newcommandx{\Adam}[1]{\todo[backgroundcolor=blue!25,bordercolor=blue,noline]{A:#1}}

\begin{document}

\begin{titlepage}       \vspace{10pt} \hfill 

\vspace{20mm}

\begin{center}

{\large \bf  Comments on Trace Anomaly Matching}

\vspace{30pt}

Adam Schwimmer$^{a}~$ and Stefan Theisen$^{b}$ 
\\[6mm]

{\small
{\it ${}^a$ Weizmann Institute of Science, Rehovot 76100, Israel}\\[2mm]
{\it ${}^b$Max-Planck-Institut f\"ur Gravitationsphysik, Albert-Einstein-Institut,\\ 
14476, Golm, Germany}
}

\vspace{20pt}

\end{center}

\centerline{{\bf{Abstract}}}
The structure of type A and B trace anomalies is reanalyzed in terms of the universal behaviour of dimension $-2$ invariant amplitudes.
Based on it a general argument for trace anomaly matching between the unbroken and broken phases of a CFT is given.
The   structure of moduli trace anomalies and their transformations under source reparametrizations is discussed  in detail.
\vspace*{5mm}
\noindent

\vspace{15pt}

\end{titlepage}

\tableofcontents{}
\vspace{1cm}
\bigskip\hrule

\section{Introduction}

Trace anomalies \cite{DDI,Duff} have rather special properties compared with the other QFT anomalies.
While chiral anomalies can be described generally in a topological framework,  
which allows their understanding independent of the group (continuous or discrete) or 
the order of the symmetry (zero form or higher form),
trace anomalies do not have such a topological description. This difference is 
related to trace anomalies being ``real", i.e. appearing as a real term in the 
Euclidean generating functional in counterdistinction  to the chiral anomalies which 
appear as a phase   
(of course in Minkowski metric all terms being phases the distinction is not there). 
As a consequence, while the 't Hooft matching for chiral anomalies, i.e. the constancy 
of the anomaly along the RG flow, follows from the topological invariants being 
rigid, such an argument for matching is not available for trace anomalies.
Nevertheless it is believed that trace anomalies are matched between the unbroken 
and spontaneously broken phases of a given CFT \cite{ST}. For this matching one should rely on 
the detailed analytic structure of the anomalous correlators.

The diffeomorphism and Weyl symmetry Ward identities obeyed by 
connected correlators  of primary operators 
have the same form
in the unbroken and broken phases. This  follows from the fact that they are derived from operatorial relations
which are the same in the two phases, evaluated on a Poincar\'e invariant vacuum, while the transformation of the vacuum under 
dilations and special conformal transformations is not used. Moreover in both phases the general analytic structure of the 
invariant amplitudes is the same. As a consequence the cohomological structures
of the generating functional are the same in the two phases. Therefore the same 
local functionals of the sources can appear as anomalies.
We mean by ``matching" simply
that the normalizations of the anomalies are the same in the two phases of a given theory.
Since the functional dependence on the momentum invariants of the correlators
is completely different in the two phases, ``anomaly matching", if valid, gives non-trivial 
constraints on e.g.  the structure and normalization  of the amplitudes  in the broken phase involving the dilaton.  

Generically the spectrum of the broken phase is massive.  The mass scale is provided by the 
non-zero vacuum expectation value of a scalar primary operator of positive dimension
which causes the spontaneous breaking of conformal symmetry to Poincar\'e symmetry. 
There could be a decoupled massless subsector which still preserves conformal invariance. 
If present it should 
be factored out in the anomaly matching. 
In the rest the only generic massless field present is the dilaton, the Goldstone boson corresponding to  
the broken Weyl symmetry. Since trace anomalies require  the contributions of massless fields as intermediate 
states in correlators, the anomaly matching fixes certain couplings of the dilaton.
These couplings are normalized by the difference between the anomalies in the unbroken 
phase and the conformal sector of the broken phase, if present.
    
Proving  anomaly matching for CFTs is not trivial \cite{ST}.
The distinction between the two types of trace anomalies \cite{DS}
(``type A" and ``type B") played an important role.
Type A anomalies have an analytic structure very similar to zero form, 
continuous group chiral anomalies. One could identify a dimension $-2$ invariant 
amplitude which, for special kinematic configurations where there is only one independent invariant $q^2$, has 
a $\frac {a}{q^2}$ dependence.  The coefficient of this power gives the normalization of the type A anomaly. The existence of 
this is again a  consequence of the conformal Ward identities  
which are also valid in the spontaneously broken phase.
In addition one needs again the usual 
requirements of analyticity which are believed to be valid also in the broken phase and therefore
the $\frac {1}{q^2} $ behaviour is also there.
Using the general property that in the limit where we rescale the momentum to infinity 
  the amplitudes in the broken phase should match those 
in the unbroken phase in the same limit, the coefficients of $\frac {1}{q^2}$
singularities should match since they originate from amplitudes which match. 
Moreover the dilaton contributes exactly to this amplitude and therefore 
the dilaton couplings are constrained by the type A anomaly coefficient calculated 
in the unbroken phase.

For type B anomalies the situation is considerably more involved. We  start by reviewing 
the  procedure for finding the normalization of  type B anomalies\cite{DDI}. 
Type B anomalies 
appear generically in correlators of integer dimensional primary operators with the 
energy-momentum tensor. One particular case involves correlators of just energy-momentum 
tensors in even dimensions.
In the cohomological analysis  type B anomalies are characterized by an anomaly 
density which is Weyl invariant and the anomaly does not vanish for 
$x$-independent Weyl parameter $\sigma$. For the standard example let us consider the 
Weyl anomaly in $d=4$ for a CFT coupled to a background metric $g_{\mu\nu}$.
Then the type B anomaly is
\be\label{eq1}    
\delta_\s W=c\,\int d^4 x\,\s\,\sqrt{g}\,C^2
\ee
where $W$ is the generating functional for connected correlation functions of the energy-momentum tensor, 
$C^2$ is the square of the Weyl tensor and $c$ the anomaly coefficient.
For $x$-independent $\sigma$,   \eqref{eq1} is also the variation of the correlators 
under dilations and therefore the anomaly is directly related to
the only possible UV counterterm
\be\label{eq2}
\bar c\,\log\Lambda^2\int d^4 x\,\sqrt{g}\,C^2
\ee
corresponding to a logarithmic UV divergence which is  possible  for integral dimension primaries in a CFT.
The correlators are  no longer  invariant under dilations since after the subtraction of the counterterm \eqref{eq2} 
the finite correlator contains terms with logarithmic dependence on the invariants.
Therefore in the unbroken phase the anomaly coefficient $c$ can be identified by looking at the variation under dilations of  a logarithmic term 
in the appropriate correlator, e.g.  the two-point function, whose $\Lambda$ dependence follows from  the second term 
in the expansion of \eqref{eq2} around flat space.
\be\label{eq3}
\int d^4 x\langle T_{\mu\nu}(x)\,T_{\rho\s}(0)\rangle e^{i\,p\cdot x}=\frac{4}{3}{\bar c}\,\log p^2/\Lambda^2
\Pi_{\mu\nu,\rho\s}(p)
\equiv \Gamma^{(2)}_{\mu\nu,\rho\s}(p)
\ee
where $\Pi_{\mu\nu,\rho\s}$ is the unique tensor structure which is both conserved and traceless and satisfies 
the symmetry conditions which follow from Bose symmetry of the two-point function. Its explicit form 
is given in \eqref{TT}. 
Calculating the variation under dilations of \eqref{eq3} and comparing with the second variation of \eqref{eq1} around flat 
space gives $c=2\,\bar c$.
 
In the broken phase, since the broken vacuum is not dilation invariant, the relation between Weyl transformations 
and dilations breaks down and the previous argument cannot be used.
The high momentum behaviour of the correlator of two energy-momentum tensors
is still given by the same $\bar c$ as in the unbroken phase, but  we cannot relate it 
directly to the normalization of the possible Weyl anomaly.
 
In order to match type B anomalies we re-examine the above set-up and we arrive at a different way 
to extract the anomaly normalizations from universal features of the correlators. This new way is more general 
and can be applied uniformly for all trace anomalies and also gives an alternative and more rigorous 
way for proving the matching of type A anomalies. 
The general procedure will be to analyze the Ward identities following from diffeomorphism 
and Weyl invariance after the correlators are decomposed in invariant
amplitudes. We will treat  from the beginning  the diffeomorphism 
Ward identities as non-anomalous and in the 
Weyl Ward identity we will introduce
the anomalous terms with the structure 
 prescribed by the cohomological analysis with a free normalization. 
The combined identities   relate  the anomaly to relations between dimension $-2$ amplitudes. Instead of  trying to isolate power-like behaviour
in one invariant when the other invariants are sent  to potentially singular points,   
we link the anomaly normalization to the special, universal   
behaviour of certain amplitudes when one invariant is taken to infinity the others being generic. 
Then the anomaly matching between the 
two phases follows as  a consequence of the equality of the respective amplitudes in the deep Euclidean limit.
Equivalently the high invariant behaviour is equivalent to the validity of sum rules normalized to the anomalies.
The sum rule is generically the integral over a discontinuity of an amplitude, 
one invariant being integrated  while the other two are kept at generic values.
In the broken phase another parameter which is kept fixed is the spontaneous breaking scale $v$ 
and the sum rule is valid for the whole range from $v=0$, the unbroken phase, to $v=\infty$, the IR limit of the broken phase.

The steps in this analysis are:
\begin{itemize}
\item[a)]
In the unbroken phase the logarithmically divergent amplitudes give the normalization of the anomaly 
through their relation to dilations as outlined above, but by themselves they are not anomalous. In the above example 
the logarithmically divergent amplitudes in the two- and three-point functions obey  non-anomalous 
relations as evidenced by the counterterm \eqref{eq2}, which is invariant both under diffeomorphisms and Weyl transformations.
Therefore the amplitudes having UV divergences can be eliminated from the anomaly analysis.

\item[b)]
Using the non-anomalous diffeomorphism Ward identities in the Weyl Ward identities, 
one obtains identities which involve only dimension $-2$ amplitudes. 
These identities relate the behaviour of the amplitudes when a particular  kinematical invariant on which 
it depends goes to infinity to the anomaly normalizations in both phases.

\item[c)]
 In addition one obtains  non-anomalous Ward identities which 
relate the dimension $-2$ amplitudes to cut-off independent expressions 
which are derived from the two-point function.
In the unbroken phase this determines the high invariant behaviour of the respective
 amplitudes in terms of the two-point function and when used
in  b)  relate the anomaly to the normalization of the two-point function,
replacing the usual argument.

\item[d)]
The dimension $-2$ amplitudes have the same deep Euclidean limit in the unbroken and broken phases.
Using this fact for the combinations of amplitudes appearing  in b),   one establishes the equality of the anomalies 
in the two phases,  i.e. ``anomaly matching".

\item[e)]
Once the existence and normalization of the anomalies in the broken phase are known, 
the constraints on the dilaton couplings follow from the known Weyl transformation of the dilaton.   

\item[f)] 
The anomaly equation obeyed by the dimension $-2$ amplitudes and their known high momentum behaviour 
implies sum rules for their discontinuities, normalized  by the anomaly.
In the IR limit of  the broken phase the sum rules are dominated by the dilaton contribution and the couplings 
of the dilaton can be determined. 

\end{itemize}

After these steps we arrive at a characteristic Ward identity summarising the anomaly structure for both 
type A and B trace anomalies (and also the perturbative chiral anomalies).
For the three-point function relevant for anomalies in $d=4$ one has
\be\label{eq1000}
s_1\,E_1(s_1,s_2,s_3)+s_2\,E_2(s_2,s_3,s_1)+s_3\,E_3(s_3,s_1,s_2)=ct
\ee
where $s_i\equiv p_{i}^{2}$ are the kinematical invariants ($p_i$ are the three external momenta),
$E_i$ are dimension $-2$  amplitudes and $ct$ is a constant which characterizes the strength of the anomaly, i.e. $a$ or $c$.
The basic Ward identity \eqref{eq1000} can be translated into two equivalent,  universal characterizations of the anomaly:
\be\label{eq1001}
E_i\xrightarrow{s_i\to\infty}{}\frac{ct}{s_i}+{\cal O}\left(\frac{s_j,s_k}{s_i^2}[\log s_i]^p\right)
\ee
and
\be\label{eq1002}
-\frac{1}{\pi}\int d s_i\, {\rm Im}_i E_i(s_i,s_j,s_k)=ct
\ee
where the imaginary part is obtained from the discontinuity with respect to the $s_i$ invariant while the other two invariants $s_j,s_k$ are kept fixed.
 
From comparing \eqref{eq1000} in the deep Euclidean limit in the unbroken and broken phases, 
one reaches the conclusion that \eqref{eq1000} and therefore \eqref{eq1001}  and \eqref{eq1002} are valid with 
the same value of the anomaly $ct$ also in the broken phase. This  gives the most general statement about
``anomaly matching". In the broken phase the functional dependence is completely different and the various amplitudes depend 
on the breaking scale $v$, but the anomaly equations are independent of $v$. In particular the relations 
are valid also for $v=\infty$,  the IR regime of the broken phase, where they impose constraints on the dilaton couplings.
In addition  for type B anomalies the normalization of the anomaly
obtained form the three-point function as outlined above is related to the two-point correlator in a universal 
fashion involving again only dimension $-2 $ amplitudes.

In Section 2 we study in detail the above scenario for the 
simplest type B anomaly in $d=4$, which  involves scalar primary operators of dimension $+2$. We 
will refer to this as the $\Delta=2$ model.
The relatively simple kinematics allows us to follow in detail the steps outlined above.
Whenever the explicit Ward identities realize a step  described above  we give the general, abstract  form of the equations /arguments 
which are valid for all anomalies. This section contains therefore our general results with the simplest explicit realization.

In Section 3 the special features related to anomalies of higher dimensional 
primaries are studied, in particular for conformal moduli in $d=4$.
We show how the general arguments can be applied also in these cases by mapping the 
high dimension amplitudes to combinations of dimension $-2$ amplitudes.  
 
In Section 4 we study in detail the analytic structure of the type B anomaly 
in the correlators of just  
energy-momentum tensors in $d=4$ and apply the general procedure for the matching of both  type A and type B anomalies.

We verify different aspects of the anomaly structure  discussed in the main text by a Feynman diagram calculation 
in a free model in Appendix A. The calculations have general validity for the $\Delta=2$ model,
since different CFT with dimension $+2$ primaries have the same analytic structure for the relevant two and three-point
correlators differing possibly just by their normalizations. 
 
A simple  explicit model for the spontaneous breaking of conformal symmetry is discussed in Appendix B.
The various general features of the anomaly structure in the broken phase are verified  and the role of 
the dilaton as an effective description of the anomaly difference for massive flows
is also exemplified.

\section{Detailed Analysis of the $\Delta=2$ Model}

Consider in $d=4$ a CFT which has a dimension two primary scalar operator ${\cal O}$. An explicit 
realization of such a model is a 
free massless scalar $\phi$ for which  
${\cal O}=\phi^2$. In Appendices A and B we present several explicit checks of our general results 
for this simple model, but our arguments will be independent of the actual realization. 

We couple the operator to a source 
$J$ which transforms under a Weyl transformation as
\be\label{eq4}
\delta_\s J=-2\,\s\,J
\ee
while the metric transforms as 
\be\label{eq4a}
\delta_\s g_{\mu\nu}=2\,\s\,g_{\mu\nu}
\ee 
The cohomological analysis gives a type~B anomaly in the Weyl transformation of the generating functional of 
connected correlation functions of the energy-momentum tensor $T_{\mu\nu}$ and ${\cal O}$, 
\be\label{eq5}
\delta_\s W=c\,\int d^4 x\,\s\,\sqrt{g}\, J^2
\ee
while diffeomorphisms are not  anomalous.
 
Even though the theory is conformal, there are logarithmic UV divergences in momentum space  correlators 
of integer dimensional operators, which require counterterms. In  particular for correlators of two $\Delta=2$ operators 
with any number of energy-momentum tensors the unique counterterm is
\be\label{eq6}
\bar c\,\log\Lambda^2\int d^4 x\,\sqrt{g}\,J^2
\ee
The standard  argument relates the anomaly coefficient $c$ to the normalization $\bar c$ 
by considering an  $x$-independent Weyl transformation which also represents dilations.
Then the explicit breaking of dilations due to the presence of the cut-off $\Lambda$ in the counterterm  
leads to a nonvanishing  Weyl variation, i.e. to an  anomaly \eqref{eq5}, 
as discussed in the Introduction. This fixes to $c=2\,\bar c$, as will be confirmed below. 

In the following we will discuss an alternative argument which avoids  amplitudes with  
UV divergences by using the high momentum behaviour of finite  dimension $-2$ invariant amplitudes.
The invariant amplitudes which contain UV divergences can be identified by expanding the metric dependence in the 
counterterm \eqref{eq6} in perturbations $h$ around flat space $\eta$ 
\be\label{eq7}
g^{\mu\nu}=\eta^{\mu\nu}+h^{\mu\nu}
\ee
In the unbroken phase the two-point function which one obtains by expanding 
the generating functional to order $J^2$
is completely determined by the dimension of ${\cal O}$  and  
in momentum space has the expression
\be\label{eq8}
\Gamma^{(2)}(p^2)\equiv\langle {\cal O}(-p)\,{\cal O}(p)\rangle=-2\,\bar c\,\log p^2/\Lambda^2
\ee
In the renormalized correlator the cut-off $\Lambda$ is replaced by a 
finite scale but we will continue using the cut-off as a scale. 
Expanding \eqref{eq6} a logarithmically divergent term with the same 
normalization will appear also in 
the correlator of two operators ${\cal O}$ and one energy-momentum tensor. 
Expanding $\sqrt{g}=1-\frac{1}{2}\eta_{\mu\nu}h^{\mu\nu}$ (using \eqref{eq7}) one finds that the 
divergence will be in a structure proportional to $\eta_{\mu\nu}$.

We will now study  this correlator by decomposing it into invariant amplitudes 
in momentum space as
\be\label{eq9}
\ba 
\Gamma^{(3)}(q,k_1,k_2)&\equiv \langle T_{\mu\nu}(-q)\,{\cal O}(k_1)\,{\cal O}(k_2)\rangle\\
\noalign{\vskip.2cm}
&=A\,\eta_{\mu\nu}+B\, q_\mu q_\nu+C\,(q_\mu\, r_\nu+q_\nu\, r_\mu)+D\, r_\mu r_\nu
\ea
\ee
where $A,B,C,D$ depend on the three Lorentz invariants $q^2,k_1^2,k_2^2$, 
with $q_{\mu}$ and $k_{1\mu},k_{2\mu}$ the four momenta carried by the 
energy-momentum tensor and the two scalar operators, respectively. In \eqref{eq9} we have defined  
\begin{subequations}\label{eq10}
\be
r_\mu=k_{1\mu}-k_{2\mu}
\ee
\hbox{and by momentum conservation one has}
\be
q_\mu=k_{1\mu}+k_{2\mu}
\ee
\end{subequations}
We remark that the amplitudes $A,B,D$ are symmetric and the amplitude $C$ is 
antisymmetric under the interchange of the momenta $k_1, k_2$.
The amplitude $A$ has dimension $0$ while the amplitudes $B,C,D$ have dimension $-2$ and 
are therefore finite, i.e. independent of the cut-off.

We now study the Ward identities which relate $\Gamma^{(3)}$ to $\Gamma^{(2)}$.
Invariance  under infinitesimal diffeomorphisms $x^\mu\to x^\mu-\xi^\mu(x)$ under which 
$g_{\mu\nu}$ and $J$ transform as 
\be\label{eq11}
\ba
\delta_\xi g_{\mu\nu}&=\nabla_\mu\xi_\nu+\nabla_\nu\xi_\mu\\
\noalign{\vskip.2cm}
\delta_\xi J&=\xi^\mu\p_\mu J
\ea
\ee
applied to the expansions of the generating functional leads to\footnote{More details about the 
derivation of the Ward identities will be given in Section 4.}
\be\label{eq12}
q^\mu\Gamma_{\mu\nu}^{(3)}(q,k_1,k_2)=k_{1\nu}\,\Gamma^{(2)}(k_2^2)
+k_{2\nu}\Gamma^{(2)}(k_1^2)
\ee
Weyl invariance, defined as the variations with parameter $\sigma(x)$ given in 
eqs.\eqref{eq4} and \eqref{eq4a},
leads to the relation
\be\label{eq14}
\eta^{\mu\u} \Gamma^{(3)}_{\m\nu}=\Gamma^{(2)}(k_1^2)+\Gamma^{(2)}(k_2^2)+2 \,c
\ee
Relations \eqref{eq12} and \eqref{eq14},  when rewritten in terms of the 
invariant amplitudes, give
\be\label{eq15}
\ba
&A+q^2 B+q\cdot r\,C=\frac{1}{2}\left[\Gamma^{(2)}(k_1^2)+\Gamma^{(2)}(k_2^2)\right]\\
&q^2\, C+q\cdot r\,D=\frac{1}{2}\left[\Gamma^{(2)}(k_2^2)-\Gamma^{(2)}(k_1^2)\right]\\
&4\, A+q^2\, B+2\,q\cdot r\, C+r^2\, D=2\left[\Gamma^{(2)}(k_1^2)+\Gamma^{(2)}(k_2^2)\right] +2\,c
\ea
\ee
where we used that the Ward identities which follow from diffeomorphism invariance are 
not anomalous 
and in the identity resulting from Weyl transformations we included the contribution of the 
anomaly obtained from the expansion of \eqref{eq5}.

From \eqref{eq15} we could replace $A$ by
\be\label{eq16}
\bar A\equiv A-\frac{1}{2}\left[\Gamma^{(2)}(k_1^2)+\Gamma^{(2)}(k_2^2)\right]
\ee
and all cut-off dependent terms disappear from the Ward identities. 
This is a consequence of the structure of the counterterm \eqref{eq6}, 
which confirms that these terms 
obey the Ward identities. 
Generically there remains a  difference between the two logarithms which does not contain
the cut-off and therefore the possible anomalies are produced by finite amplitudes.  
 More generally we can simply solve the first equation of \eqref{eq15} for $A$, replace it in the third and obtain
\begin{subequations}\label{eq17}
\be
q^2\,C+q\cdot r\, D=\frac{1}{2}\left[\Gamma^{(2)}(k_2^2)
-\Gamma^{(2)}(k_1^2)\right]\label{eq17a}
\ee
\be\label{eq17b}
-3\,q^2\, B-2\,q\!\cdot\! r\,C+r^2\, D=2\,c 
\ee
\end{subequations}
We stress that all amplitudes present in \eqref{eq17} have dimension $-2$ 
and the contribution from the two-point function is  also finite, keeping the information about its overall normalization.
The appearance of the Ward identities with the structure of \eqref{eq17} is generic and we will now discuss 
their properties and role for the matching in the general setting.

Equation \eqref{eq17b} is a particular instance of the general type of equations \eqref{eq1000}
\be\label{eq100}
s_1\,E_1(s_1,s_2,s_3)+s_2\,E_2(s_2,s_3,s_1)+s_3\,E_3(s_3,s_1,s_2)=2\,c
\ee
where $s_1=q^2,\,s_2=k_1^2$ and $s_3=k_2^2$ are the three kinematical invariants and  
$E_i$ are the dimension $-2$  amplitudes
\be\label{Es}
E_1=-3\,B-D\,,\qquad E_2=2(D-C)\,,\qquad E_3=2(D+C)
\ee
 
The amplitudes with dimension $-2$  obey unsubtracted dispersion relations in any of the 
invariants when the other two invariants are kept fixed at generic values.
We choose for each amplitude the invariant with the same index since this is the dependence 
constrained by the anomaly equation \eqref{eq100}, i.e.
\be\label{eq101}
E_i(s_i,s_j,s_k)=\frac{1}{\pi}\int dx_{i}\frac{{\rm Im}_i E_i(x_i,s_j,s_k)}{x_i-s_i}
\ee
where Im$_{i}$ indicates  ${1\over 2 i}\times$ the discontinuity in the variable $s_i$, while the other two invariants are kept fixed.
We remark that \eqref{eq101} also contains the information about  the analytic structure in the variables which are kept fixed, 
after doing appropriate analytic continuations.
In a CFT the support of the integral is between $0$ and $\infty$ for the $x_i $ variable.  
We choose $s_j,s_k$ to be  real negative in order to have a nonsingular discontinuity.  

The large $s$ behaviour of a dimension $-2$ invariant amplitude in a CFT is generically ${1\over s} [\log(s)]^p$  
for any of the invariants, where the scale of $s$ in the $\log$ is given by the invariants which are kept fixed.  
If the amplitude however satisfies  \eqref{eq101}, the behaviour is more restricted:
taking a discontinuity in $s_i$ of \eqref{eq100} we obtain
\be\label{eq102}
s_i\,{\rm Im}_i E_i+s_j\,{\rm Im}_i E_j+s_k\,{\rm Im}_i E_k=0
\ee
which implies
\begin{subequations}\label{eq103}
\be
{\rm Im}_i E_i\xrightarrow[s_i\to\infty]{}\frac{1}{s_i^2}[\log s_i]^p+\dots\label{eq103a}
\ee
\be
E_i\xrightarrow[s_i\to\infty]{}\frac{2\,c}{s_i}+{\cal O}\left(\frac{s_j,s_k}{s_i^2}[\log s_i]^p\right)\label{eq103b}
\ee
\end{subequations}
Then $s_i$ can be taken outside the dispersion relation, 
and comparing with \eqref{eq103} 
we obtain the sum rules for each of the invariant amplitudes:
\be\label{eq104}
-\frac{1}{\pi}\int d s_i\, {\rm Im}_i E_i(s_i,s_j,s_k)=2\,c
\ee
Therefore if an invariant amplitude $E_i$ which appears in an anomaly equation of the form \eqref{eq100} 
obeys any of the equivalent
universal relations \eqref{eq103} or \eqref{eq104}, the parameter $c$ gives directly  the anomaly coefficient.
This special structure  \eqref{eq100} of the Ward identity for dimension $-2$ amplitudes, relating it to the anomaly,  
is generic and common also to the type A trace anomaly and even to chiral anomalies. Once it is obeyed
the high invariant behaviour of the amplitudes \eqref{eq103} or, equivalently, the sum rules \eqref{eq104} follow.

What makes type B anomalies special is the relation of the anomaly  coefficient $c$  to the two-point function.
For type B anomalies typically there is a diffeomorphism Ward identity with a cut-off independent contribution 
of the two-point function  which fixes the special high invariant contribution of the form \eqref{eq103} 
recovering this way the relation between the anomaly normalization and the two-point function, as we 
now show explicitly for the $\Delta=2$ model. 

For this model we analyze eq.\eqref{eq17} at 
\be\label{eq18}
x\equiv k_1^2-k_2^2=0
\ee
which is not a singular point.
The amplitudes depend on $q^2$ and on
\be\label{eq19}
k^2\equiv k_1^2=k_2^2
\ee
Taking a derivative with respect to $x$ at $x=0$ of \eqref{eq17a} and evaluating 
\eqref{eq17b} at $x=0$, we obtain
\begin{subequations}\label{eq20}
\be
q^2\,\bar C+D=\frac{\bar c}{k^2}\label{eq20a}
\ee
\be
q^2\big(-3\,B-D)+4\,k^2\,D=2\,c \label{eq20b}
\ee
\end{subequations}
where we used that the amplitude $C$ is odd in $x$ and we defined a dimension $-4$ 
amplitude $\bar C$ by
\be\label{eq21}
\bar C(q^2,k^2)\equiv\frac{\p C}{\p x}(q^2,k^2,x=0)
\ee
Now we can use the high $k^2$ behaviour for $D$ extracted from \eqref{eq20a}:
\begin{subequations}\label{eq105}
\be\label{105a}
D\xrightarrow[k^2\to\infty]{}\frac{\bar c}{k^2}+{\cal O}\left(\frac{q^2}{k^4}[\log k^2]^p\right)
\ee
and compare it with the relevant equation following from \eqref{eq103}: 
\be\label{105b}
D\xrightarrow[k^2\to\infty]{}\frac{c}{2\,k^2}+{\cal O}\left(\frac{q^2}{k^4}[\log k^2]^p\right)
\ee
\end{subequations}
leading to the equality $c=2\,\bar c$.
This argument,  which relates the normalizations of the type B anomaly and of the two-point function 
in the unbroken phase is general for all the type B anomalies: besides the equation \eqref{eq17b} there 
is always  an equation generalizing \eqref{eq17a} which relates the high invariant behaviour
of the amplitude to the normalization of the two-point function. Comparing the two we get the desired relation between the anomaly 
and the two-point function normalizations without using UV divergent amplitudes. 

Another special feature of type B anomalies is the appearance of effective IR poles, reflecting 
the role of the two-point correlator in the Ward identity.
We will demonstrate this in the concrete setting for the $\Delta=2$ model.
We proved  in the unbroken phase the relation between the anomaly normalization and the two-point function 
using  the special kinematic configurations  $k_{1}^{2}=k_{2}^{2}\equiv k^2$ and $q^2$.
If we assume in addition the validity of dispersion relations in the ``diagonal variable" $k^2$, by the argument
following \eqref{eq100}, we obtain the sum rule 
\be\label{eq204}
-\frac{1}{\pi}\int dk^2\, {\rm Im}_{k^2} D(k^2,q^2)=\frac{c}{2}
\ee
If this sum rule is valid also  for $q^2=0$ then, since there is no scale left for ${\rm Im}_{k^2}  D$, we conclude
\be\label{eq205}
{\rm Im}_{k^2}  D(k^2,0)=-\frac{\pi\,c}{2}\, \delta(k^2) \qquad \hbox{and}\qquad     D(k^2,0)=\frac {c}{2\,k^2}
\ee
This functional dependence shows the presence of an effective zero-mass pole.   
The mechanism for its appearance is simple:
for  $q^2=0$, from eq.\eqref{eq17a} we obtain
\be
{\rm Im}_{k^2}D(k^2,0)=\lim_{k_1^2\to k_2^2}
\frac{{\rm Im}_{k_2^2}\Gamma^{(2)}(k_2^2)
-{\rm Im}_{k_1^2}\Gamma^{(2)}(k_1^2)}{2(k_1^2-k_2^2)}
\ee
Since\footnote{We use the definition of  the logarithm as a real analytic function on the first Riemann sheet 
with the branch cut on the positive real axis.}
 ${\rm Im}_{k^2}\log k^2=-\pi\,\theta(k^2)$, the result is as in Figure 1,
%\be
\begin{figure}[htb]
\centering
\begin{tikzpicture}[scale=1.5]
\draw[style=thick](-2,0)--(-1,0);
\draw[style=thin](-1,0)--(1,0);
\draw[style=thick](1,0)--(2,0);
\draw[style=thick](-1,1.5)--(1,1.5);
\draw[style=dashed](-1,0)--(-1,1.5);
\draw[style=dashed](1,0)--(1,1.5);
\draw[style={->,thin}](0,0)--(0,1.8) ;
\draw node [shift={(0.0,3.0)}] {${\rm Im}_{k^2}D$}; 
\draw node [shift={(0.3,1.9,0)}] {$\frac{c \pi}{2 x}$};
\draw node [shift={(-1.5,-.3)}] {$-\frac{x}{2}$}; 
\draw node [shift={(+1.5,-.3)}] {$+\frac{x}{2}$}; 
\end{tikzpicture}
\caption{}
\end{figure}
%\ee
which is a regularized $\delta$-function and therefore, using also \eqref{eq8} the limit is
\be
{\rm Im}_{k^2}D(k^2,0)\to -\bar c\,\pi\,\delta(k^2)
\ee
and comparing with \eqref{eq205} gives again $c=2\,\bar c$. 
We remark that the appearance of the $\delta$-function is the result 
of a ``collision"  between the branch points in $k_1^{2}$ and $k_2^{2}$.
This effective pole is specific to type B anomalies and is different  from the generic presence of poles following 
from the anomaly sum rules which represent the collapse of ordinary branch cuts in certain limits. 
In particular it is not matched in the broken phase where the analytic structure of the two-point correlator is completely different. 

Once the high invariant  behaviour and sum rules for dimension $-2$ amplitudes \eqref{eq103} and \eqref{eq104} are valid  for type B, 
we could discuss in general the matching for all trace anomalies.
We start our discussion of the anomaly matching with a summary of the structure of the spontaneously broken phase. 
Let us assume that there is another Poincar\'e invariant vacuum on which a nonzero dimensional
scalar primary operator gets a vacuum expectation value. In such a situation the conformal symmetry is 
spontaneously broken and a mass scale $v$, introduced through the vacuum expectation value,
the order parameter of the broken phase, is introduce in the theory.
There are several general characteristics of the broken phase which we will use:

\noindent
a) Following from Goldstone's theorem a zero mass scalar, the dilaton exists. 
The dilaton $\mathcal{D}$ 
has a linear coupling to the energy-momentum tensor with a dimensional 
strength $f$  related to $v$
\be\label{eq200}
\langle 0|T_{\mu\nu}|{\Sigma(q)}\rangle=f\, q_\mu q_\nu
\ee
Formally one  has Weyl invariance also in the presence of the coupling \eqref{eq200} if we attribute to the dilaton the Weyl transformation
\be\label{eq201}
\Sigma  \rightarrow \Sigma+\sigma
\ee
where we used  the normalized dimensionless dilaton field $\Sigma\equiv \mathcal{D}/f$. In the broken phase Weyl invariance is limited  
to transformations where $\sigma(x)$ falls off to zero at large $x$. As a consequence  e.g. the relation between dilations 
and Weyl invariance with $x$-independent $\sigma$ is not anymore valid.  

\noindent
b) Since the operatorial relations of the CFT are not changed and the derivation 
of the diffeomorphism and 
Weyl Ward identities used only the Poincar\'e invariance of the vacuum, all the Ward identities 
we used in the unbroken phase remain valid. Also the analyticity properties of the invariant 
amplitudes remain valid. Anomalies can appear only as real parts in Ward identities corresponding 
to the same type of anomaly functionals as in the unbroken phase.  
Therefore the basic Ward identity for the dimension $-2$ amplitudes $E_{i}^{B}$ in the broken phase, characterized by the 
mass scale $v$, is valid but with  an a priori different  normalization of the anomaly $c^{B}$: 
\be\label{eq400}
\ba
s_1\,E_1^{B}(s_1,s_2,s_3,v^2)+s_2\,E_2^{B}(s_1,s_2,s_3,v^2)+s_3\,E_3^{B}(s_1,s_2,s_3,v^2)=2\,c^{B}
\ea
\ee
The anomalies match if $c^B=c$. 
From its high momentum analysis we conclude, as in the unbroken phase,
\be\label{eq401}
E_i^{B}\xrightarrow[s_i\to\infty]{}\frac{2\,c^{B}}{s_i}+{\cal O}\left(\frac{s_j,s_k,v^2}{s_i^2}[\log s_i]^p\right)
\ee 
and
\be\label{eq402}
-\frac{1}{\pi}\int d s_i\, {\rm Im}_i E_i^{B}(s_i,s_j,s_k,v)=2\,c^{B}
\ee
where we made explicit the possible dependence of the discontinuities on the breaking scale $v$.

\noindent
c) For a  given correlator the deep Euclidean limit of the amplitude in 
the broken phase coincides 
with the limit in the unbroken phase. 
Denoting invariant amplitudes in the two phases as ${\cal A}^B$ and ${\cal A}$, respectively
we have
\be\label{eq31}
\lim_{\lambda\to\infty}\frac{{\cal A}(\lambda\, q_1^2,\lambda\, q_2^2,\dots)}
{{\cal A}_B(\lambda\, q_1^2,\lambda\, q_2^2,\dots)}=1
\ee
the deviations being of order $1/\lambda$ or $v^2/q_i^{2}$ when the 
invariants are taken to $\infty$.
This is simply a consequence of the fact that the OPE of the operators are not changed in the 
broken phase and therefore the UV structure of the correlators remains the same even though 
for finite values of the invariants the structure of the amplitudes changes 
in the broken phase, the spectrum of the theory being generically massive, etc.

Anomaly matching is now an immediate consequence. Consider the combination of dimension $-2$ invariant amplitudes 
which appear in the ``anomaly equations" \eqref{eq100} and \eqref{eq400}, respectively,   
for a configuration in the deep Euclidean limit for generic 
(i.e. avoiding special points like $s_j=0$) configurations. We have
\be\label{eq403}
\lim_{\lambda\to\infty} \frac{ \lambda s_1\,E_1(\lambda s_1,\lambda s_2,\lambda s_3)+\lambda s_2\,E_2(\lambda s_1,\lambda s_2,\lambda s_3)
+\lambda s_3\,E_3(\lambda s_1,\lambda s_2,\lambda s_3)}
{\lambda s_1\,E_1^{B}(\lambda s_1,\lambda s_2,\lambda s_3)+\lambda s_2\,E_2^{B}(\lambda s_1,\lambda s_2,\lambda s_3)
+\lambda s_3\,E_3^{B}(\lambda s_1,\lambda s_2,\lambda s_3)}
=\frac{c}{c^{B}}=1
\ee
where we used that $c,c^{B}$ do not depend on the invariants or on the breaking scale $v$.

Once the equality of the anomalies is established,  it follows that asymptotic values of $E_i$ for taking $s_i$ to $\infty$ 
and the sum rules \eqref{eq104} and \eqref{eq402}  also match.
The sum rules \eqref{eq402}, now normalized to, $c$ are valid  for the whole range of $v$ including $v=0$, the unbroken phase. 
 
We can relate the sum rules in special limits to particular contributions in the two phases. Consider the sum rule for the discontinuity in $q^2$ 
of the amplitude $-3B-D$, which we denoted  by $E_1$ in the unbroken phase:
\be\label{eq900}
-\frac{1}{\pi}\int d q^2\, {\rm Im}_{q^2} E_1(q^2,k_{1}^2,k_{2}^{2})=2\,c
\ee  
At $ k_{1}^{2}=0$ or $k_{2}^{2}=0$ the amplitude is singular since one has a branch point. One can, however, approach the configuration 
$k_{1}^{2}=k_{2}^{2}\equiv k^2=0$ as a limit in $k^2$ approaching $0$. Since \eqref{eq900} holds also in the limit and the integrand has dimension $-2$
this implies
\be\label{eq910}
{\rm Im}_{q^2} E_1(q^2,0,0)=-2\,c\,\pi\, \delta(q^2)
\ee
This pole-like discontinuity, which is  reached in a very special way due to the singularity of the limit, gives a universal characterization of the anomaly. 
As shown above the characterizations through  the high invariant limit of the amplitude \eqref{eq103} or equivalently the sum rules \eqref{eq104}
are much more general and mathematically unambiguous.
  
In the broken phase generically the sum rule is saturated with massive states. If however we go to the  deep IR limit, i.e.  $v \rightarrow \infty$  
when all the masses are sent to $\infty$ then also in the broken phase the saturation will  be  due only to massless states and \eqref{eq910}
must be valid. Generically in the broken phase there is a sector which preserves conformal invariance which therefore could contribute to $c$.
Outside this sector the only generic massless state is the dilaton  whose coupling in the $q^2$-channel can produce the $\delta (q^2)$ dependence normalized 
to the coupling of the dilaton to the rest of the diagram. Therefore the anomaly matching will constrain the dilaton couplings requiring their proportionality 
to the difference between the anomaly in the unbroken phase and the anomaly of the conformal sector in the broken phase.
 
In the deep IR limit of the broken phase the dilaton reproduces completely the anomaly.
We recall the implementation  of this general relation.
Assume that in the presence of the external sources, the metric $g_{\mu\nu}$ and the sources $J$ coupled to the additional primaries,  one has an anomaly:
\be\label{eq201}
\delta_{\sigma} W(g_{\mu\nu},J)=\int d^4x\, \sigma\mathcal{A}(g_{\mu\nu},J)
\ee
where $W$ is the generating functional and  $\mathcal{A}$ is the local anomaly functional containing  the normalization mentioned above.
Then the dilaton effective action $S(g_{\mu\nu},J,\Sigma)$, whose variation reproduces the anomaly \eqref{eq201}, is
\be\label{eq202}
S(g_{\mu\nu},J,\Sigma)=-\int_{0}^{1} dt \int d^4x \,\Sigma \,\mathcal{A}(g_{\mu\nu}^{-t\,\Sigma}, J^{-t\,\Sigma}) +\Psi(g_{\mu\nu}^{-\Sigma}, J^{-\Sigma})
\ee
where $g_{\mu\nu}^{-t\,\Sigma}, J^{-t\,\Sigma}$ are the sources transformed by a Weyl parameter $\sigma=-t\,\Sigma$  and $\Psi$ an 
arbitrary diffeomorphism invariant functional, contributing a Weyl invariant term.
For the $\Delta=2$ model the dilaton effective action  is
\be\label{eq203}
S(g_{\mu\nu},J,\Sigma)=-c\,\int\,d^4 x\sqrt{g}\, \Sigma\, J^2 +\Psi\big(g_{\mu\nu}\exp{(-2\Sigma)} , J\exp{(2\,\Sigma)}\big)
\ee
The second term is invariant under diffeomorphism and Weyl transformations. 
The first term represents the "dilaton coupling" to the two operators ${\cal O}\, {\cal O}$. In principle using its analytic and 
covariance properties it can be separated from the general off-shell $\langle\Sigma\,{\cal O}\,{\cal O}\rangle$ correlator.
In the IR limit of the broken phase when  the effective action is expanded in powers of all the momenta it  is singled out by 
being the only ``ultralocal" term which contains  $J^2$ and survives when all  momenta are zero. 
More generally, in the deep IR the dilaton effective action is given by a polynomial expansion in momenta and the
dilaton coupling will always be defined such that it corresponds to the lowest independent  
terms in the momentum expansion around zero. 

\section{Analysis of the Moduli Problem}

Consider, again in $d=4$, a dimension four scalar primary which has the special property 
that it does not have a $\beta$-function, i.e. in particular  its structure constant vanishes. 
Such a primary, called ``modulus" in the following, will have nevertheless a 
Type B anomaly induced by its two-point function.\footnote{Various aspects of this type of anomalies 
were studied in refs. \cite{Gomis,Pomoni1,Pomoni2,Pomoni3}.} 
In the first part of this section we will study the 
anomaly structure of a CFT with one modulus. This will be generalized in the second part 
to the case of several moduli. 

\subsection{The Anomaly Structure} 

The high dimension of the modulus 
compared with the $\Delta=2$ model of the previous section, 
produces new features which we will analyze. Coupling the 
operator to a source $J$, which is Weyl invariant, the anomaly is 
\be\label{eq1p}
\delta_\s W=c\,\int d^4 x\,\s\,\sqrt{g}\,J\,\Delta_4 J
\ee
where $\Delta_4$ is the Fradkin-Tseytlin-Paneitz-Riegert (FTPR) operator \cite{FT,Paneitz,Riegert} 
with the special property that it transforms homogeneously under 
Weyl transformations with weight $-4$, i.e. $\Delta_4\to e^{-4\,\s}\Delta_4$. Its explicit form will be given 
in \eqref{eq601}. 

The anomaly reflects the logarithmic divergence in correlators of two moduli with 
any number of energy-momentum tensors. The corresponding diffeomorphism and Weyl 
invariant counterterm is
\be\label{eq2p}
\bar c\,\log\Lambda^2\,\int d^4x\,\sqrt{g}\,J\,\Delta_4 J
\ee
The two-point function is
\be\label{eq3p}
\langle{\cal O}(p)\,{\cal O}(-p)\rangle\equiv \Gamma^{(2}(p)=-2\,\bar c\,(p^2)^2\,\log p^2/\Lambda^2
\ee
As for the previous case, we want to understand the cut-off independent  characterization of the anomaly 
as it appears in the correlator of two moduli and one energy-momentum tensor. 
Since the modulus operator ${\cal O}$ is a Lorentz scalar, the decomposition in invariant 
amplitudes is formally identical to the one in the previous section, \eqref{eq9}, but now the invariant 
amplitudes have positive dimensions: $+4$ for the $A$ amplitude 
and $+2$ for the $B,C,D$ amplitudes.
The derivation of the Ward identities is similar; the diffeo identities are identical to the 
first two equations of \eqref{eq15}, while the Weyl equation, the third equation in
\eqref{eq15}, does not have a right hand side besides the anomaly since the source $J$ is invariant 
under a Weyl transformation.

The amplitudes have UV divergences. Since we want to preserve conformal invariance, the 
normalization conditions corresponding to power divergences are put to zero. 
Therefore the $A$ and $B,C,D$
amplitudes obey triply, respectively doubly subtracted dispersion relations. 
In order to deal with finite amplitudes we use the fact that the logarithmically divergent 
contributions obey the Ward identities, since the counterterm is invariant under both 
diffeomorphisms and Weyl transformations. 
We can then shift the amplitudes by terms containing logarithms and these shifts will remove 
the two-point contributions which contain the cut-off since the counterterm fixes completely the form of the logarithmic divergence.
We list the shifts which follow from the structure of the counterterm \eqref{eq2p}:
\be\label{eq4p}
\ba
A&\to A+\frac{\bar c}{6}\big(q^4-q^2(k_1^2+k_2^2)-6\, k_1^2\,k_2^2\big)
\left(\log k_1^2/\Lambda^2+\log k_2^2/\Lambda^2\right)\\
B&\to B-\frac{\bar c}{6}\big(q^2-k_1^2-k_2^2)
\left(\log k_1^2/\Lambda^2+\log k_2^2/\Lambda^2\right)\\
C&\to C-\frac{\bar c}{2}\big(k_1^2-k_2^2\big)
\left(\log k_1^2/\Lambda^2+\log k_2^2/\Lambda^2\right)\\
D&\to D+\frac{\bar c}{2}\big(q^2+k_1^2+k_2^2\big)
\left(\log k_1^2/\Lambda^2+\log k_2^2/\Lambda^2\right)
\ea
\ee
After these shifts the Ward Identities for the finite shifted amplitudes  have the form 
\begin{subequations}\label{eq5p}
\be
A+q^2\, B+q\cdot r\,C=\frac{\bar c}{2}\big(k_1^4-k_2^4)\big(\log k_2^2 -\log k_1^2\big)\label{eq5ap}
\ee
\be
q^2\,C+q\cdot r\,D=\frac{\bar c}{2}\big(k_1^4+k_2^4\big)\big(\log k_1^2-\log k_2^2\big)
\label{eq5bp}
\ee
\smallskip
\be
4\, A+q^2\,B+2\,q\cdot r\,C+r^2\,D = c \, \big((k_{1}^{2})^2 +(k_{2}^{2})^2\big)\label{eq5cp}
\ee
\end{subequations}
where we assumed that the Weyl Ward identity can be anomalous and we used the form of the anomaly \eqref{eq1p}
expanded around the flat metric $\eta_{\mu\nu}$.

Using \eqref{eq5ap} we  re-express \eqref{eq5cp} in terms of the dimension $2$ amplitudes
\be\label{eq7p}
-q^2\big(3\,B+D\big)-2\,q\cdot r\,\,C+2\big(k_1^2+k_2^2\big)D
+2\,\bar c\big(k_1^4-k_2^4\big)\big(\log k_1^2-\log k_2^2\big)=\frac{c}{4}\big((k_{1}^{2})^2 +(k_{2}^{2})^2 \big)
\ee
We can absorb the additional $k$ dependent term in the l.h.s. in a redefinition of $C$:
\be\label{eq301}
C\equiv\bar C -\bar c\,\big( k_{1}^{2} +k_{2}^{2}\big)\big(\log k_{1}^{2}-\log k_{2}^{2}\big)
\ee
Now \eqref{eq301} has the form 
\be\label{eq302}
s_1\,E_1(s_1,s_2,s_3)+s_2\,E_2(s_1,s_2,s_3)+s_3\,E_3(s_1,s_2,s_3)=Q(s_1, s_2,s_3)
\ee
which generalizes \eqref{eq100}. We pause again to discuss in general the properties  of this positive dimensional anomaly structure.
In \eqref{eq301} the amplitudes $E_i$ have dimension $N$, while $Q$, which contains the normalization of the anomaly,
is a homogenous polynomial in $s_1,s_2,s_3$ of dimension 
$N+2$. To keep the discussion general, we take $N$ to be a positive even integer or $0$.
The amplitudes $E_i$ have the special feature that they do not contain the cut-off scale, 
i.e. they are UV convergent in spite of their non-negative dimension.
This means that they have convergent dispersion relations in any of the $s$ variables. 
This is possible only if the amplitude as an analytic function has all its singularities in a dimension $-2$ function and 
the overall dimension is made up by integer powers of the $s$ variables.  
Therefore the amplitudes $E_i$ have the special form
\be\label{eq302}
E_i(s_1,s_2,s_3)= \Sigma _k P_k(s_1,s_2,s_3)\tilde E_{i}^{(k)}(s_1,s_2,s_3)
\ee
where $P_k$ are monomials formed from the $s$-variables of total dimension $N+2$ and $E_{i}^{(k)}$ are 
dimension $-2$ analytic functions. The summation is over all monomials which are compatible with the total dimension $N$. 
The dispersion relations for $E_i$ will be convergent: the discontinuity is coming from $\tilde E_{i}^{(k)}$ 
multiplied by the monomial  and in the dispersion relation itself the monomial is simply taken outside the 
integral if it does not involve the integration variable $s_i$; if $s_i$ is part of the monomial it  
provides ``free" subtractions at $s_i=0$, getting also outside the integral.
 
Using \eqref{eq302}  we can repeat our discussion following \eqref{eq100} to determine the 
special features of the dimension $-2$ amplitudes $\tilde E_{i}^{(k)}(s_1,s_2,s_3)$ related 
to the presence of the anomaly polynomial $Q$ in the r.h.s.
Taking an  asymptotic expansion in each of the variables $s_i$ and equating 
the corresponding terms on the two sides of the equation one obtains the equivalent relations
\be\label{eq303}
\tilde E_{i}^{(k)}\xrightarrow[s_i\to\infty]{}\frac{f_{i}^{k}(c)}{s_i}+{\cal O}\left(\frac{s_j,s_k}{s_i^2}[\log s_i]^p\right)
\ee
and 
\be\label{eq304}
-\frac{1}{\pi}\int d x_i {\rm Im}_i \tilde E_{i}^{(k)}(x_i,s_j,s_k)=f_{i}^{k}(c)
\ee
where $f_{i}^{k}(c)$ are pure numbers depending on the anomaly polynomial $Q$. 
In particular some of these coefficients could be zero if the appropriate term does not appear 
in $Q$. The leading terms in the asymptotic expansion give relations 
as specified by \eqref{eq304} while non-leading ones have generically sums of terms.
Saturating  \eqref{eq304} with $\delta$-function type discontinuities at special configurations 
when only one invariant is left is generically problematic also in this case.
For type B anomalies at least one of the expressions appearing in \eqref{eq303} is given 
by a diffeomorphism Ward identity involving the two-point function  and then comparing it 
with \eqref{eq303} we get the desired relation between the anomaly in the unbroken phase 
and the normalization of the two-point function.
 
We return now to the moduli anomaly. To simplify again the argument we choose the nonsingular 
kinematical configuration $k_{1}^{2}=k_{2}^2\equiv k^2$.
For the $D$ amplitude the relevant expansion is
\be\label{eq305}
D(q^2,k^2)=(k^2)^2 \tilde D(q^2,k^2) +....
\ee
and then clearly the behaviour of $\tilde D$ is similar to the behaviour of $D$ for the $\Delta=2$ model 
of the previous section. From the asymptotic expansion of  \eqref{eq5bp}  in $k^2$ we obtain
\be\label{eq306}
\tilde D \rightarrow \frac{\bar c}{k^2}
\ee
and using the expansion in \eqref{eq7p} we find $c=2\,\bar c$ from the matching with the $(k^2)^2$ in the anomaly polynomial. 

Clearly after the  expansion of the positive dimension amplitudes 
in terms of dimension $-2$ amplitudes the moduli problem (and a similar one for the anomalies 
of dimension $3$ scalar operators) are mapped to the $\Delta=2$ case.
The anomaly matching follows from an argument similar to the one used in Section 2 for the $\Delta=2$ model:
one considers again the ratio of \eqref{eq302} in the unbroken and broken phases in the deep Euclidean limit.
Equating the ratio of the anomaly polynomials in the same limit the equality of the normalizations follows.
In particular for a single modulus the dilaton effective action is:
\be\label{eq307}
S(g_{\mu\nu},J,\Sigma) =-c\,\int d^4 x\,\Sigma\,\sqrt{g}\,J\,\Delta_4 J +\Psi(g_{\mu\nu }\exp{-2\Sigma},J)
\ee
where $\Psi$ is an arbitrary diffeomorphism invariant functional. 
The dilaton coupling is  defined  by the normalization of the unique local term with four derivatives  in the expansion of the effective action.

We comment that a similar analysis can be performed for the type B anomaly generated by a 
dimension $+3$ scalar primary coupled to a dimension $-1$ source $J$ where
\be\label{eq800}
\delta_\s W=c\,\int d^4 x\,\s\,\sqrt{g}\,J\,\Delta_2 J \qquad\qquad \Delta_2=\square  -\tfrac{1}{6} R
\ee
\subsection{The role of source reparametrizations}

We consider now  moduli in $d=4$, i.e. dimension $4$ scalar primaries ${\cal O}_i,  i=1,...,N,$
which have the special property that the structure  constants of any three moduli (including the same modulus) vanish.
This prevents the appearance of  higher than linear terms  of $\log{\Lambda^2}$ in the three-point 
correlator and therefore a vanishing of the $\beta$ function in lowest order. We assume that the higher 
order constraints leading to the exact vanishing of the $\beta$ function are also fulfilled.
Then the moduli can be added to the action with finite coefficients 
and conformality is preserved.
We diagonalize the two-point correlators of the moduli which is always possible in a unitary CFT and we 
normalize the operators such that after diagonalization  the two-point function 
is proportional to the unit matrix.

We couple the moduli to sources $J^i$ via\footnote{In this section we use $\g$ for the space-time 
metric and reserve $g$ for the Zamolodchikov metric.}
\be\label{eq600}
\int d^4 x\, \sqrt{\g}\, \sum_{i=1}^{N}J^{i}\, {\cal O}_{i}
\ee
The sources are inert under Weyl transformations and 
as a consequence one  can redefine them 
through local functions without interfering with their Weyl transformation properties.
We will discuss the meaning and role of these transformations for 
type B anomalies.
Since different type B anomalies can mix under the transformations we need a complete list of these anomalies.

In order to simplify the analysis, we will use the one-to-one relation between type B anomalies and 
logarithmic counterterms which is valid in the unbroken phase and we will classify the  
logarithmic counterterms. Since counterterms preserve the diffeomorphism 
and Weyl symmetries, they are constructed from local scalar integrands which transform homogeneously  with weight $-4$ 
under Weyl transformations.
We list these local expressions for a single source $J$ when $J$ is acted upon by derivative operators.  
\be\label{eq601}
I_1(J)\equiv  \Delta_4 J\equiv \Big(\Box^2 +\frac{1}{3}(\nabla^\mu R)\,\nabla_\mu+2\, R^{\mu\nu}\nabla_\mu\nabla_\nu
-\frac{2}{3}R\,\square\Big)J
\ee
where $\Delta_4$ is the FTPR operator \cite{FT,Paneitz,Riegert}.
\be\label{eq602}
I_2(J)\equiv 
\square J\,\square J+2\,\nabla^\mu\nabla^\nu J\,\nabla_\mu\nabla_\nu J
+4\,\nabla^\mu J\,\nabla_\mu\square J
+4\left(R^{\mu\nu}-\frac{1}{6}g^{\mu\nu}R\right)\nabla_\mu J\,\nabla_\nu J
\ee

\be\label{eq603}
I_3(J)\equiv \Box J\nabla^{\mu}J\nabla_{\mu}J + 2\nabla^{\mu}\nabla^{\nu}J\nabla_{\mu}J\nabla_{\nu}J
\ee

\be\label{eq604}
I_4(J)\equiv \nabla^{\mu}J\nabla_{\mu}J \nabla^{\nu}J\nabla_{\nu}J 
\ee

\noindent
The counterterms contain the integrated expressions 
\be\label{eq605}
C_k \equiv \int d^4x \sqrt{\gamma}\,I_k(J)
\ee
Since $I_1,I_3$ are total derivatives the corresponding $C_1,C_3$ vanish. Of course we expect these 
vanishings since in the unbroken phase there is no expectation value for ${\cal O}$ and the correlator of three ${\cal O}$ 
also vanishes. Therefore we will get just the counterterms $C_2,C_4$.
We can get form the densities $I_1,I_3$ non-vanishing counterterms if we multiply them with additional powers of $J$.  
These counterterms are, however,  not independent. For example
\be\label{eq606}
\int d^4x \sqrt{\gamma}\, J\, I_1(J)=-C_2 
\ee
and
\be\label{eq607}
\int d^4 x \sqrt{\gamma}\, J\,I_3(J)= -C_4
\ee
The counterterms produce type B anomalies by using the same integrands in the variation of the effective action\footnote{This relation 
to type B anomalies which, by definition, have a Weyl invariant anomaly density, is the reason why above we listed only Weyl invariant densities.} 
\be\label{608}
\delta_{\sigma}W=c_k \int d^4x \,\sigma(x)\,\sqrt{\gamma}\, I_k(J)  \qquad\qquad  k=2,4
\ee
where $c_k$ gives the normalization of the anomaly.

The equivalence of counterterms 
of the type discussed above, which was  based on integration by parts, 
can produce additional terms when derivatives act on $\sigma$. These expressions, being Weyl invariant 
and vanishing for $x$-independent $\s$,  represent therefore,
if cohomologically nontrivial, possible type A anomalies. It is an interesting question if type A anomalies for moduli can be 
realized physically,  but we limit our discussion just to type B and therefore we ignore the possible type A anomalies 
which may appear in the equivalence relations.

When we have more than one source, any combination of sources in the expressions above would produce a priori independent anomalies. 
The counterterms /anomalies \eqref{eq602},\eqref{eq604} represent just the simplest terms in infinite 
families of moduli anomalies. Consider, as an example starting from \eqref{eq602}, the family
of  anomalies
\be\label{eq24p}
\ba
c_{\{k\}ij}\int &d^4 x\,\s\,\sqrt{\g}\,J^{k_1}\dots J^{k_n}\,\Big(
\square J^i\,\square J^j+2\,\nabla^\mu\nabla^\nu J^i\,\nabla_\mu\nabla_\nu J^j\\
&\qquad+4\,\nabla^\mu J^i\,\nabla_\mu\square J^j
+4\big(R^{\mu\nu}-\tfrac{1}{6}g^{\mu\nu}R\big)\nabla_\mu J^i\,\nabla_\nu J^j\Big)
\ea
\ee
with a priori independent universal, i.e. scheme independent normalizations  $c_{\{k\}ij}$.
Obviously these expressions are still Weyl invariant and they represent new possible anomalies. 

They correspond to 
single logarithmic divergences in the correlators of $k+2$ moduli with theory dependent normalizations. 
In expressions \eqref{eq24p} summation over all $(k+2)!$ permutations of the $k+2$ indices of 
$c_{\{k\}ij}$ are understood, irrespective of whether they are all different or not. This ensures Bose 
symmetry of the moduli correlators which are derived from them. This also applies to all the 
following expressions. 

 The reduction of counterterms built from the $I_1,I_3$ structures to the  counterterms of type $C_2,C_4$ is valid 
in the general situation when we have an arbitrary number of sources and the structures  can be multiplied  by arbitrary 
products of sources. This is a consequence of the complete symmetrization on the sources which is valid for all our expressions.
Indeed let us consider a typical identity for a single source used in the integration by parts when proving the 
equivalence of the counterterms:
\be\label{eq609}
\Box J^p=p J^{p-1}\Box J+ p(p-1)J^{p-2}\,\nabla_{\mu}J\nabla^{\mu}J
\ee
The same relation is valid if we have complete symmetrization of $p$ sources,  i.e.
\be\label{eq610}
\square(J_{i_{1}}...J_{i_{p}})_S=p (\square J_{i_{1}}...J_{i_{p}})_S +p(p-1)(\nabla_{\mu}J_{i_{1}}\nabla^{\mu}J_{i_{2}}...J_{i_{p}})_S
\ee
 where we put an index $S$ to remind us that the expressions are completely symmetrized.
It follows that we can repeat the integration by parts manipulations for many sources by simply doing the calculation 
for a single source and in the final result replace the appropriate expressions for many sources, completely symmetrized.
We obtain that generically an expression based on $I_1$ gives us linear combinations of expressions based on 
$I_2$ and $I_4$ while an expression based on $I_3$ gives just expressions of the $I_4$ type.
 
A convenient way to treat all the independent anomalies corresponding to $I_2 $ and $I_4$ is 
to sum over the sources and then the two families of 
anomalies will be characterized by two families of functions $h_{ij}(J)$ and $t_{ijkl}(J)$, respectively:
\be\label{eq25p}
\ba
\int d^4 x\,&\s\,\sqrt{\g}\, h_{ij}(J)\,\Big(\square J^i\,\square J^j+2\,\nabla^\mu\nabla^\nu J^i\,\nabla_\mu\nabla_\nu J^j\\
&\qquad\qquad+4\,\nabla^\mu J^i\,\nabla_\mu\square J^j
+4\big(R^{\mu\nu}-\tfrac{1}{6}g^{\mu\nu}R\big)\nabla_\mu J^i\,\nabla_\nu J^j\Big)
\ea
\ee
\be\label{eq611}
\int d^4 x\,\s\,\sqrt{\g}\,t_{ijkl}(J)\nabla^{\mu}J^{i}\nabla_{\mu}J^{j} \nabla^{\nu}J^{k}\nabla_{\nu}J^{l}
\ee 
The coefficients $c_{\{k\}ij}$ are recovered from $h_{ij}(J)$ and $t_{ijkl}(J)$ by a Taylor expansion around $J_i=0$ for $i=1,...,N$.

These two  sets  of functions contain all the information about the Type B 
anomalies of moduli in $d=4$. All the expressions are completely symmetrized in the sources.
A partial constraint following from this is that $h_{ij}$ and $t_{ijkl}$ are symmetric under interchange of $i$ with $ j$ 
and of $k$ with $l$.

Consider now possible reparametrizations of the sources
\be\label{eq26p}
J^k=f^k(J')
\ee
where $f^k$ are  Taylor expandable around $J'^{i}=0$, invertible as a power series  and start with a normalized linear term, i.e. 
\be\label{eq608}
f^k(J')=J'^{k}+ {\cal O}((J')^2)
\ee
Obviously the sources $J'^{i}$ are also inert under Weyl transformations. 
The change of variables induces  reparametrizations  in the generating functional, which now will depend on $J'$,   
and therefore a reparametrization of the coefficients of the terms which  contain  single logarithms
which are at the origin of the anomalies. To follow  the same convention  after the change of variables we 
do a complete summation over permutations of the $J'$ variables.
Since the possible anomalies with $J'^{i}$ as sources are characterized by the same basis of space-time integrands, now written 
in terms of $J'^{i}$, the new coefficient functions can be read off the explicit form of the anomaly.
As the simplest example we consider the anomaly \eqref{eq611} which was first discussed in \cite{Osborn}. 

Doing the reparametrization of the anomaly the transformation of the 
coefficient function is
\be\label{eq609}
t'_{ijkl}(J')=t_{mnpr}(J(J'))\frac{\partial J^m}{\partial J'^{i}}\frac{\partial J^n}{\partial J'^{j}}\frac{\partial J^p}{\partial J'^{k}}\frac{\partial J^r}{\partial J'^{l}}
\ee
A similar procedure will give the transformation rules  for $h'_{ij}(J')$,  which will involve a linear combination of 
$h_{ij}(J)$ and $t_{ijkl}(J)$. Clearly the functional dependence of the anomaly functions is not 
fully physical since changing the sources through reparametrization does not change  the physical 
meaning while changing the functional form.  We stress that there is no a priori constraint on the transformation 
properties of the anomaly functions: they follow from the explicit form of the anomalies we choose as 
a basis  by simply doing the change of variables on the anomaly formulae.
 
We need  to understand the relation between the reparametrizations and the correlators of the moduli. 
The original sources $J^i$ were coupled to the moduli operators ${\cal O}_j$ through the coupling \eqref{eq600}.
The correlators of the moduli were defined  by the Taylor expansion of the generating functional, i.e.  
the coefficient  of $J^{k_1}(x_1)...J^{k_n}(x_n)$ in its expansion gives the correlator
\be\label{eq28p}
\langle{\cal O}_{k_1}(x_1)\,{\cal O}_{k_2}(x_2)\cdots{\cal O}_{k_n}(x_n)\rangle
\ee
After the change of variables, expanding now in powers of $J'$, their coefficients 
will be given by linear combinations of the general form
\be\label{eq29p}
\ba
&\langle{\cal O}_{k_1}(x_1)\,{\cal O}_{k_2}(x_2)\cdots{\cal O}_{k_n}(x_n)\rangle
+\sum_{i\neq j}a^r_{ij}\delta(x_i-x_j)\langle {\cal O}_{k_1}(x_1)\cdots {\cal O}_{k_r}(x_j)\cdots{\cal O}_{k_n}(x_n)\rangle\\
&\qquad+\sum_{i\neq j\neq k\neq i}b^r_{ijk}\delta(x_i-x_j)\,\delta(x_i-x_k)
\langle{\cal O}_{k_1}(x_1)\cdots {\cal O}_{k_r}(x_k)\cdots{\cal O}_{k_n}(x_n)\rangle+\dots
\ea
\ee
The second term of the first line is a sum over  $(n\!-\!1)$-point functions where 
the two operators ${\cal O}_{k_i}(x_i)$ and ${\cal O}_{k_j}(x_j)$ have been replaced by $\delta(x_i-x_j) a^r_{ij}{\cal O}_{k_r}(x_j)$,  
the third line is a sum of $(n\!-\!2)$-point functions 
where three operators have been replaced by one, etc. The coefficients $a^r_{ij},\, b^r_{ijk}$ etc. are 
determined by the (derivatives of the) functions $f^r$ in the change of basis \eqref{eq26p}. 
Therefore, compared with the original correlators, the new ones have ``semilocal" contributions when expressed 
in terms of the old correlators, i.e. contributions of lower order correlators multiplied by $\delta$-functions.
The only correlators which are not changed by the reparametrization invariance are the two-point functions since  the 
one-point functions, which could have given a semilocal contribution,  are zero in the unbroken phase.
So depending on the sources, semilocal contributions could be present. We do not, however, have an intrinsic definition for
such contributions and we can identify only their relative appearance between two sets of correlators which correspond
to two choices of sources, which are related by a reparametrization. We therefore conclude that all choices of sources  
which are related by reparametrizations are equivalent and contain the same universal  
information.  Therefore, all dependence on the sources, including the functional form of the anomaly functions,
should be taken modulo reparametrizations in order to obtain the regularization independent universal information about the respective CFT.

Once this is understood we  could obtain additional information about the anomaly functions.
Let us  consider, as an example, the three $J$ contributions in the anomalies.
Since there is no intrinsic anomaly starting with three $J$, terms of this type can appear only in the $h_{ij}$ anomaly.
In light of the previous discussion this means  that this must be true modulo reparametrizations, 
i.e. there should exist a choice of sources for which the three  
$J$ terms vanish. But in an arbitrary parametrization one could have a three $J$ term, reflecting a semilocal three-point function:

\be\label{eq610}
\delta(x-y)\langle {\cal O}(y)\,{\cal O}(z)\rangle
\ee
It follows that three $J$ contributions in the $h_{ij}$ anomaly are possible since by choosing the 
quadratic terms in the $f^k(J)$ functions we could put them to zero.

We do not have similar constraints for \eqref{eq611}. Even though the structure constants vanish for three moduli, 
one can still have an unremovable logarithm in correlators with four and more moduli:
in a block decomposition one has couplings between two moduli and other primaries which are not moduli
and a logarithm may be produced.

Finally we want to use the previous discussion to produce a basis for the two remaining independent anomalies  
which have simple transformation rules for the anomaly functions characterizing them.
We stress that this is not a logical necessity which imposes constraints on the theory
but just a convenient choice. The functions $t_{ijkl}$ for the \eqref{eq611} anomaly  already have 
simple transformation rules \eqref{eq609}, so we will concentrate on \eqref{eq25p}. 
 
The new form  should still be a functional only  of $h_{ij}$ which contains all the universal information. 
Therefore the new terms we add could depend only on $h_{ij}$. They should be Weyl invariant  in order 
that the new form continues to be a type B anomaly. We have the option to add a term based on the 
kinematical structure $I_3(J)$  which we know to be reducible to $I_4(J)$. We use this option 
normalizing the contribution to a ``connection" derived from $h_{ij}$. In addition we can add a term with 
the form of \eqref{eq611} but again with a normalization dependent on $h_{ij}$. Therefore at the end these two 
modifications amount to a redefinition of $t_{ijkl}$ by an additive $h_{ij}$  dependent term.
We arrive therefore at the new  form of \eqref{eq25p}
\be\label{eq30p}
\ba
\hat {\cal A}=&\int\sqrt{\gamma}\,\sigma\,g_{ij}
\Big\lbrace \hat\square J^i\,\hat\square J^j+2\,\hat\nabla^\mu\hat\nabla^\nu J^i\,
\hat\nabla_\mu\hat\nabla_\nu J^j
+4\,\nabla^\mu J^i\,\hat\nabla_\mu\hat\square J^j\\
&\qquad\qquad+4\left(R^{\mu\nu}-\frac{1}{6}g^{\mu\nu}R\right)\nabla_\mu J^i\,
\nabla_\nu J^j\Big\rbrace 
\ea
\ee
where 
\be
\ba
\hat\nabla_\mu\hat\nabla_\nu J^i&=\nabla_\mu\nabla_\nu J^i+\Gamma^i_{kl}\,\p_\mu J^k\,\p_\nu J^l\\
\hat\nabla_\mu\hat\square J^i&=\nabla_\mu\hat\square j^i+\Gamma^i_{kl}\,
\nabla_\mu J^k\,\hat\square J^l
\ea
\ee
and 
\be
\Gamma^i_{jk}=\frac{1}{2}g^{im}\left(\p_k g_{mj}+\p_l g_{jm}-\p_m g_{ij}\right)
\ee
and we replaced  $h_{ij}$ with  $g_{ij}$  to stress that they have different transformation properties under reparametrizations.
Indeed if we work out the reparametrization of \eqref{eq30p} we obtain that $g_{ij}$ transforms as a symmetric tensor. 
We remark again that \eqref{eq25p} and \eqref{eq30p} are completely equivalent as far as the universal information 
they carry is concerned and are equally correct forms for the type B anomaly.
Due to the simple transformation properties of $g_{ij}$ it is easy to check the vanishing requirement for thee three $J$ contribution 
in a given ``frame": simply one can choose Riemann  normal coordinates where $g_{ij}(0)=\delta_{ij}$ and 
$\Gamma^i_{jk}(0)=0$, and then from the form of \eqref{eq30p} it is evident. 
 
The expression \eqref{eq30p} is clearly Weyl invariant, being a linear combination of the previous anomalies with 
special choices of the anomaly functions.  

We now discuss the meaning and transformation properties of the 
``Zamolodchikov metric" related to the above general discussion.
The Zamolodchikov metric is defined \cite{Zamolodchikov}
by first deforming the original CFT 
through the addition of a term 
\be\label{eq34p}
\sum_k\bar J^k\int d^4 x\,\sqrt{\g}\,{\cal O}_k
\ee
where 
$\bar J^k$ are finite deformation parameters.
Then the two-point functions of the moduli 
$\langle{\cal O}_j {\cal O}_l\rangle$
are studied in the deformed theory and their normalization, defined by a 
matrix $\bar g_{jl}({\bar J})$, gives the ``Zamolodchikov metric".
Since the analytic structure of the two-point function is completely fixed by conformal invariance,
the normalization is also the normalization of the logarithmic dependence and therefore the metric is 
closely related to the type B anomaly. We want to study the details of this relation, in particular
the covariance of the metric under a reparametrization of the deformation parameters $\bar J^k$.

Though the generating functional's dependence on the sources is defined as an expansion 
in $J^k$ around $J^k=0$, it is believed to have a finite radius 
of convergence and therefore can be expanded   
also around $\bar J^k$,  giving the correlators of the deformed theory.

To discuss this concretely, let us choose the ``covariant scheme" \eqref{eq30p}
and expand $J^k(x)$ as:
\be\label{eq35p}
J^k(x)=\bar J^k+\tilde J^k(x)
\ee
Then Taylor expanding in $\tilde J$ and keeping only quadratic terms, 
we obtain for the anomaly
\be\label{eq36p}
\ba
{\cal A}&=\int d^4 x\,\s\sqrt{\gamma}\,g_{ij}(\bar J)\Big[\square\tilde J^j\,\square\tilde J^l
+2\,\nabla^\mu\nabla^\nu\tilde J^j\,\nabla_\mu\nabla_\nu\tilde J^l\\
&\qquad+4\,\nabla^\mu\tilde J^j\,\nabla_\mu\square\tilde  J^l
+4\big(R^{\mu\nu}-\tfrac{1}{6}g^{\mu\nu}R\big)\nabla_\mu\tilde J^j\,\nabla_\nu \tilde J^l
\Big]
\ea
\ee
This expression can be completed with the additional terms in $\tilde J^k$ to make it a type B  
anomaly in the deformed theory at generic $J^{k}$,  
but they are not needed for our argument. The lowest term above represents the metric in $\tilde J^k$ at $\bar J^k$, which
is the normalization of the two-point function of the moduli without insertions of the energy-momentum tensor,  i.e. the Zamolodchikov metric.
 
Therefore one can immediately identify $g_{ij}({\bar J})$ with the Zamolodchikov metric.
It contains all the universal information about the infinite class of 
type B anomalies  defined in the undeformed theory, i.e. the coefficients of the single logarithms in correlators of
any number of moduli operators.
The covariance properties of the Zamolodchikov metric are simply inherited from those of the anomaly metric.
Even though the metric reflects the two-point function at the deformed point it reflects the infinite summation of all higher order correlators.
Different parametrizations of the deformed point  contain the scheme dependence as semi-local contributions could contribute  and produce 
the transformation of the metric. The semilocal contributions do not have an intrinsic (universal) meaning, unless 
some higher symmetry is introduced.

The above argument misses an important aspect of the Zamolodchikov metric as characterizing also  the global features 
of the moduli space. It assumes that the “path” between the ``perturbative expansion point $J=0$'' and the finite point $\bar J$ is unique. 
This is not generically true: one can have ``holonomies” on the moduli space related to its global properties. 
The anomaly approach being  intrinsically perturbative misses the information about this 
structure and it is a very interesting problem to try to find such a global information in the anomalies.

Finally the above identification gives us a simple argument for the matching of the $g_{ij(J)}$ anomaly normalizations.
The Zamolodchikov metric gives the normalization of the two-point function for the deformed theory at $\bar J^k$. 
As we argued in Section 3 this normalization is matched through the relation to the three-point function to 
the anomaly in the broken phase. The matching occurs for a give scheme for $J$ and it is covariant under a reparametrization. 
 
Similar considerations can be made for the second anomaly \eqref{eq611} in the deformed theory. 
Its normalization for the lowest term, the logarithmic term in the correlator of four moduli,
defines a Zamolodchikov-Osborn tensor  which is given by
$t_{ijlm}(\bar J^j)$. Its covariance properties under repametrizations of $\bar J^k$ are again given by \eqref{eq609}.

\section{Energy-Momentum Tensor Three-Point Function} 

We now turn to the case in which Weyl anomalies were originally discussed, the 
three-point function of the energy-momentum tensor in $d=4$, 
\be\label{Gamma3}
\Gamma^{(3)}_{\mu\nu,\rho\s,\a\b}(k_1,k_2,k_3)\equiv
\langle T_{\mu\nu}(k_1)\,T_{\rho\s}(k_2)\,T_{\\a\b}(k_3)\rangle\qquad\hbox{with}\qquad  k_1+k_2+k_3=0
\ee 
It is the ${\cal O}(h^3)$ term in the expansion of the effective action around Minkowski space.   
It exhibits both Type A and Type B anomalies \cite{OP,OE,Cappelli,Skenderis}. Here we will be mainly concerned with the latter. Compared 
to the correlation functions which we discussed in the previous sections, the one discussed 
here is far more involved due to the more complicated tensor structure. 

The correlation function \eqref{Gamma3} has dimension four and, in addition to the symmetries implied 
by the symmetry of $T_{\mu\nu}$, it has $S_3$ Bose symmetry under permutation of the three pairs of indices.
As already discussed  in detail in \cite{Cappelli}, there are 137 possible tensor structures, each of which is multiplied 
by an invariant amplitude which is a function of the three independent kinematical invariants 
$k_1^2,k_2^2,k_3^2$. The tensor indices can be carried by the three momenta and the Minkowski metric.
As the total dimension $\Gamma^{(3)}$ is four, the amplitudes which multiply tensor structures 
where all six indices are carried by momenta, have dimension $-2$, those where four indices are carried 
by momenta have dimension 0, those where two indices are carried by momenta have dimension 
$+2$ and, finally, those where all indices are carried by the metric, have dimension $+4$. 
Their numbers are 27, 63, 42 and 5, respectively. Among those only the 27 dimension $-2$ 
amplitudes are scheme independent and therefore unambiguous and our analysis is based on them. 

In general space-time dimension the tensor structures are independent, however in integer dimensions 
there are dimension-dependent  special identities, so-called Schouten identities. 
For $d=4$, which we are interested in, there is one 
identity among the dimension zero tensor structures which is the vanishing of 
the third metric variation of $\int\sqrt{g}E_4$ (c.f. \eqref{Euler}). It vanishes as the 
integrand is a total derivatives.  

Our aim is to generalize the analysis of the $\Delta=2$ model of Section 2. More specifically,  
by choosing appropriate linear combinations we look for  
diffeomorphism and Weyl Ward identities which involve only the 
27 dimension $-2$ amplitudes. In a second step we use a diffeomorphism Ward identity 
to fix the normalization of the Type B anomaly in terms of the normalization of the 
two-point function. 

Due to Bose symmetry the invariant amplitudes come in $S_3$ orbits, where $S_3$ acts 
on the arguments. There are orbits of length 6, 3, 2 and 1. In the first case the amplitudes 
have no symmetry under exchange of any two arguments while in the last case they are totally 
symmetric. To make the $S_3$ symmetry manifest we use a basis for the tensor structures where 
the independent momenta are chosen as follows: the indices $(\mu,\nu)$ are 
carried by $k_2,k_3$, indices $(\rho,\s)$ by $k_3,k_1$ and $(\a,\b)$ by $k_1,k_2$.
Of course the indices can also be carried by the metric. A similar discussion 
can be found in  \cite{Skenderis}, which we closely followed. 

As for the final analysis of the Ward identities only the dimension $-2$ amplitudes, where all 
six tensor indices are carried by momenta, enter, we will 
only enumerate those.
We introduce the following notation for their tensor structures, e.g. 
\be
(23;13;12)=k^2_\mu k^3_\nu k^1_\rho k^3_\s k^1_\a k^2_\b
\ee
and for the invariant amplitudes
\be
A_I^{\{ijl\}}=A_I(k_i^2,k_j^2,k_l^2)
\ee
Then the seven $S_3$ orbits of the dimension $-2$ amplitudes are
%\begin{subequations}
\be
\ba
(1)\quad&A_{1}^{\{123\}} (22;11;11)+ 
A_{1}^{\{213\}} (22;11;22)+ 
A_{1}^{\{132\}} (33;11;11) \\
&\qquad\qquad+ A_{1}^{\{231\}} (22;33;22) + 
A_{1}^{\{312\}} (33;33;11)+ 
A_{1}^{\{321\}} (33;33;22)\\
\noalign{\vskip.1cm}
%\ea
%\ee
%\be
%\ba
(2)\quad&A_{2}^{\{123\}} (33;11;12) + 
A_{2}^{\{213\}} (22;33;12)k +  
A_{2}^{\{132\}} (22;13;11) \\ 
&\qquad\qquad + A_{2}^{\{231\}} (23;11;22)  +
A_{2}^{\{312\}} (33;13;22)   + 
A_{2}^{\{321\}} (23;33;11) \\
\noalign{\vskip.1cm}
%\ea
%\ee
%\be
(3)\quad&A_{3}^{\{123\}} (22;33;11)  + 
A_{3}^{\{213\}}  (33;11;22)  \qquad
A_3^{\{ijk\}}=A_3^{\{jki\}}=A_3^{\{kij\}}\\
\noalign{\vskip.1cm}
%\ee
%\be
(4)\quad &A_{4}^{\{123\}} (22;11;12)  + 
A_{4}^{\{312\}} (33;13;11)   +
A_{4}^{\{321\}} (23;33;22)  \qquad
A_4^{\{ijk\}}=A_4^{\{jik\}}\\
\noalign{\vskip.1cm}
%\ee
%\be
(5)\quad&A_{5}^{\{123\}} (23;13;12)  \qquad
A_5^{\{123\}}=A_5^{\{231\}}=A_5^{\{312\}}=A_5^{\{213\}}=A_5^{\{321\}}=A_5^{\{132\}}\\
\noalign{\vskip.1cm}
%\ee
%\be
%\ba
(6)\quad&A_{6}^{\{123\}}  (23;11;12)  + 
A_{6}^{\{213\}}  (22;13;12)   +  
A_{6}^{\{132\}}  (23;13;11)  \\
&\qquad\qquad +  A_{6}^{\{231\}}  (23;13;22)    + 
A_{6}^{\{312\}}  (33;13;12)   + 
A_{6}^{\{321\}}  (23;33;12) \\   
\noalign{\vskip.1cm}
%\ea
%\ee
%\be
(7)\quad&A_{7}^{\{123\}} (23;11;11)    + 
A_{7}^{\{231\}} (22;13;22)     +
A_{7}^{\{312\}} (33;33;12)   \qquad
A_7^{\{ijk\}}=A_7^{\{ikj\}}
\ea
\ee
%\end{subequations}

\noindent
This defines the 27 dimension $-2$ invariant amplitudes, but there are only seven independent functions of three arguments. 
In any particular CFT they are fixed. 

The Ward identities are derived as follows. We expand the generating functional for connected correlation 
functions as
\be
\ba
W&=\log\int D\phi\,e^{-S[\phi,g]}\\
&=\frac{1}{2!}\int dx\,dy\,\tilde\Gamma^{(2)}_{\mu\nu,\rho\s}(x,y)\,h^{\mu\nu}(x)\,h^{\rho\s}(y)\\
&\qquad +\frac{1}{3!}\int dx\,dy\,dz\,\tilde\Gamma^{(3)}_{\mu\nu,\rho\s,\a\b}(x,y,z)\,h^{\mu\nu}(x)\,h^{\rho\s}(y)\,h^{\a\b}(z)+\dots
\ea
\ee
where
\be
h^{\mu\nu}=g^{\mu\nu}-\eta^{\mu\nu}
\ee
We note that $\tilde\Gamma^{(n)}$ differs from the $n$-point function of the energy-momentum tensor by a factor $(-1/2)^n$. 
This follows form the definition 
\be
\langle T_{\mu\nu}\rangle=-\frac{2}{\sqrt{g}}\frac{\delta}{\delta g^{\mu\nu}}W
\ee

We are interested in the variation of $W$ under infinitesimal diffeomorphisms and Weyl transformations of the metric 
\be
\ba
\delta_\xi g_{\mu\nu}&=\nabla_\mu\xi_\nu+\nabla_\nu\xi_\mu \qquad\qquad\hbox{(diffeo.})\\
\noalign{\vskip.3cm}
\delta_\s g_{\mu\nu}&=2\,\s\,g_{\mu\nu}\qquad\qquad\qquad~~~\hbox{(Weyl)}
\ea
\ee
Expanded to first order in $h^{\mu\nu}$ these translate to
\be
\ba
\delta_\xi h^{\mu\nu}&=-\p^\mu\xi^\nu-\p^\nu\xi^\mu-h^{\mu\rho}\p_\rho\xi^\nu
-h^{\nu\rho}\p_\rho\xi^\mu+\xi^\rho\p_\rho h^{\mu\nu}\qquad\hbox{(diffeo.)}\\
\noalign{\vskip.3cm}
\qquad \delta_\s h^{\mu\nu}&=-2\,\s\,(\eta^{\mu\nu}+h^{\mu\nu})\qquad\qquad\qquad\qquad\qquad\qquad\qquad~~\hbox{(Weyl)}
\ea
\ee 
As before, we assume a regularization which preserves diffeomorphism invariance of $W$. Evaluating 
$\delta_\xi W=0$ at ${\cal O}(h^2)$ results in the diffeomorphism Ward identity, which in momentum space reads
\be\label{diffeoWI}
\ba
k_1^\mu\Gamma^{(3)}_{\mu\nu,\rho\s,\a\b}(k_1,k_2,k_3)
&=k_{1\rho}\Gamma^{(2)}_{\nu\s,\a\b}(k_3)+k_{1\s}\Gamma^{(2)}_{\rho\nu,\a\b}(k_3)
+k_{1\a}\Gamma^{(2)}_{\rho\s,\nu\b}(k_2)\\
\noalign{\vskip.2cm}
&\qquad+k_{1\b}\Gamma^{(2)}_{\rho\s,\a,\nu}(k_2)
-k_{2\nu}\Gamma^{(2)}_{\rho\s,\a\b}(k_3)-k_{3\nu}\Gamma^{(2)}_{\rho\s,\a\b}(k_2)
\ea
\ee
Due to the anomaly, $W$ is not invariant under Weyl transformations: 
\be
\delta_\s W=\int d^4 x\,\sqrt{g}\,\s\,{\cal A}(x)
\ee
In $d=4$ there are two cohomologically non-trivial CP-even solutions 
to the Wess-Zumino consistency condition, 
parametrized by theory dependent constants $a$ and $c$, and a cohomologically trivial one, parametrized by $b$, 
\be\label{WeylAnomaly}
{\cal A}=c\,C^2-a\,E_4+b\,\square R
\ee
where 
\be\label{Euler}
E_4=R^{\mu\nu\rho\s}R_{\mu\nu\rho\s}-4\, R^{\mu\nu}R_{\mu\nu}+R^2
\ee
is the Euler density and $C^2=C^{\mu\nu\rho\s}C_{\mu\nu\rho\s}$ the square of the Weyl tensor.
Only the cohomologically non-trivial solutions are true anomalies as they cannot be 
removed by addition of a local counterterm to the generating functional. 

The Weyl Ward identity in momentum space is  
\be\label{WeylWI}
\ba
\eta^{\mu\nu}\Gamma^{(3)}_{\mu\nu,\rho\s,\a\b}(k_1,k_2,k_3)=2\,\Gamma^{(2)}_{\rho\s,\a\b}(k_2)+2\,\Gamma^{(2)}_{\rho\s,\a\b}(k_3)
+{\cal A}_{\rho\s,\a\b}(k_2,k_3)
\ea
\ee
where ${\cal A}_{\rho\s,\a\b}$ is obtained from the expansion of \eqref{WeylAnomaly} to ${\cal O}(h^2)$. 

For each of the Ward identities \eqref{diffeoWI} and \eqref{WeylWI} there are two more,  where 
the divergence and trace are w.r.t. to the second and third energy-momentum tensor, respectively.
They easily follow from those given by Bose symmetry. All these Ward identities hold both in the broken 
and in the unbroken phase, a priori  with different anomaly coefficients.  

The final ingredient which we need is the two-point function $\Gamma^{(2)}_{\mu\nu,\rho\s}(k)$. 
It is not anomalous and its tensor structure is fixed by conservation and tracelessness to be that of 
the ${\cal O}(h^2)$ term in the expansion of $C^2$, as discussed in the Introduction: 
\be\label{TT}
\Gamma^{(2)}_{\mu\nu.\rho\s}(k)=\Pi_{\mu\nu,\rho\s}(k)\,f(k^2)=
\left(\pi_{\mu\nu}\,\pi_{\rho\s}-\frac{3}{2}\big(\pi_{\mu\rho}\,\pi_{\nu\s}+\pi_{\mu\s}\,\pi_{\nu\rho}\big)\right)f(k^2)
\ee
where 
\be
\pi_{\mu\nu}=k^2\,\eta_{\mu\nu}-k_\mu\,k_\nu
\ee
In the unbroken phase, 
\be\label{f2}
f(k^2)=\frac{4}{3}\bar c\,\log(k^2/\Lambda^2)
\ee
while in the broken phase $f(k^2)$ is more complicated and not known generally, except that for $k^2\gg v^2$ 
it approaches \eqref{f2}. 

The analysis of the Ward identities now proceeds in several steps, which are analogous to the 
ones which we followed in Section 2. Due to the large number of tensor structures and 
invariant amplitudes and the fact that their range of dimension is from $+4$ to $-2$,  it is 
considerably more involved and we will skip most of the straightforward but tedious details.\footnote{For which we 
used the xAct Mathematica package \cite{xAct}.}  

In the first step we insert the expansion of $\Gamma^{(3)}_{\mu\nu,\rho\s,\a\b}$ in invariant amplitudes 
into the (non-anomalous) diffeomorphism Ward identities. Separating the resulting tensor structures\footnote{Here it is 
advantageous to convert to a basis which involves only two of the momenta, e.g. 
$k_1$ and $k_2$.} leads to a large 
number of homogeneous and  inhomogeneous linear relations between the invariant amplitudes. The coefficients are 
homogeneous polynomials
in the three kinematical invariants and the inhomogeneities, if present, are 
$f(k_i)$ multiplied by a non-negative power of $k_i^2$. 
By taking linear combinations we obtain relations which involve only dimension $-2$ amplitudes and $f(k_i^2)$. The 
simplest such relation which we find and which we will use later, is
\be\label{diffeoWI1}
\ba
\big(s_1\!-\!s_2\!-\!s_3\big)A_{1}^{\{213\}}+\big(s_3\!+\!s_1\!-\!s_2\big)A_{1}^{\{123\}}+\big(s_2\!-\!s_1\big)A_{4}^{\{123\}}=4\big(f(s_1)\!-\!f(s_2)\big)
\ea
\ee  
plus two others which are related by Bose symmetry. 
Here, as before,  
\be
s_1=k_1^2\,,\qquad s_2=k_2^2\,,\qquad s_3=k_3^2
\ee
This identity is satisfied in the unbroken and in the broken phase, where in the former $f(k^2)$ is 
given by \eqref{f2}. Note that in these relations the UV cut-off $\Lambda$ cancels, as required by the fact the 
all amplitudes in these relations are of dimension $-2$ and therefore finite, i.e. cut-off independent. 

We now turn to the Weyl Ward identities. Inserting the expansion of $\Gamma^{(3)}$ in invariant 
amplitudes leads to new inhomogeneous linear relations between them, where the inhomogeneities 
now contain $f(k_i^2)$ and the anomaly coefficients $a$ and $c$ and $b$. Again all coefficients are 
simple homogeneous polynomials of the kinematical invariants. 

As in the analysis of Section 2, adding to them appropriate linear combinations of the diffeomorphisms Ward identities, 
we obtain (anomalous) relations which involve only dimension $-2$ amplitudes.  The simplest 
such relation which, furthermore, does not contain $f(k_i^2)$ and only the type B Weyl anomaly 
coefficient $c$, is of the general type of \eqref{eq1000}, 
\begin{subequations}\label{WeylWI1}
\be
s_1\,E_1+s_2\, E_2+s_3\,E_3=\frac{8}{3}c
\ee
with
\be
\ba
E_1&=A_1^{\{213\}} + A_2^{\{132\}} - A_2^{\{231\}} - \frac{1}{2}A_4^{\{123\}} + A_6^{\{123\}} - A_6^{\{213\}} - A_7^{\{123\}} + A_7^{\{231\}}\\
E_2&=A_1^{\{123\}} - A_2^{\{132\}} + A_2^{\{231\}} - \frac{1}{2}A_4^{\{123\}} - A_6^{\{123\}} + A_6^{\{213\}} + A_7^{\{123\}} - A_7^{\{231\}}\\
E_3&=-A_1^{\{123\}} - A_1^{\{213\}} + A_2^{\{132\}} + A_2^{\{231\}} -\frac{1}{2} A_4^{\{123\}} - A_7^{\{123\}} - A_7^{\{231\}}
\ea
\ee
\end{subequations}  
and again two more related by Bose symmetry. 
We will show below that the anomaly $c$ is fixed by $\bar c$, the normalization of the two-point function. 

We also find anomalous  Weyl Ward identities between the dimension $-2$ amplitudes which involve only the 
type A anomaly coefficient $a$, e.g. 
\begin{subequations}\label{WeylWI2}
\be
s_1\,E_1+s_2\, E_2+s_3\,E_3=-16\,a
\ee
with
\be
\ba
E_1&=2 A_1^{\{123\}} + 4 A_1^{\{213\}} + 2 A_1^{\{132\}} + 10 A_1^{\{231\}} + 2 A_1^{\{312\}} + 2 A_1^{\{321\}} 
+ A_2^{\{123\}} - 9 A_2^{\{213\}}\\
\noalign{\vskip.1cm}
&\quad -  2 A_2^{\{132\}} - 2 A_2^{\{312\}} - 2 A_2^{\{ 321\}}+ 2 A_3^{\{123\}} - 4 A_3^{\{213\}} - 3 A_4^{\{123\}} - 2 A_4^{\{312\}} + 
 3 A_4^{\{321\}}\\
 \noalign{\vskip.1cm}
 &\qquad - A_5^{\{123\}} + A_6^{\{123\}} + 3 A_6^{\{213\}} + 2 A_6^{\{132\}} - A_6^{\{312\}} + A_6^{\{123\}} 
 - 2 A_7^{\{123\}} + 2 A_7^{\{231\}} + 7 A_7^{\{312\}}\\
 \noalign{\vskip.3cm}
E_2&=-2 A_1^{\{213\}} + 6 A_1^{\{231\}} + A_2^{\{123\}} - 9 A_2^{\{213\}} + 2 A_2^{\{231\}} + 4 A_2^{\{312\}} - 2 A_3^{\{213\}} 
+ A_4^{\{123\}} - 3 A_4^{\{321\}} \\
 \noalign{\vskip.1cm}
&\quad- A_5^{\{123\}} - A_6^{\{123\}} + A_6^{\{213\}} - 4 A_6^{\{231\}} + A_6^{\{312\}} + 9 A_6^{\{231\}} 
+ 4 A_7^{\{231\}} - 9 A_7^{\{312\}}\\
 \noalign{\vskip.3cm}
E_3&=-4 A_1^{\{213\}} - 10 A_1^{\{231\}} - 4 A_1^{\{321\}} - A_2^{\{123\}} + 5 A_2^{\{213\}} + 4 A_2^{\{231\}} - 2 A_2^{\{312\}} 
- 4 A_3^{\{213\}} - A_4^{\{123\}} \\
 \noalign{\vskip.1cm}
&\quad+  7 A_4^{\{321\}} - A_5^{\{123\}} + A_6^{\{123\}} + A_6^{\{213\}} + 2 A_6^{\{231\}} 
 + A_6^{\{312\}} - 5 A_6^{\{231\}} - 2 A_7^{\{231\}} + 5 A_7^{\{312\}}
\ea
\ee
\end{subequations}

We now analyze the Ward identities. We start with  \eqref{diffeoWI1}. At $s_2=s_1$ it is satisfied identically and contains 
no information. A non-trivial relation is obtained if we first take the derivative w.r.t. $s_1$ before setting $s_2=s_1$ and then 
taking the limit   $s_1\to\infty$ while keeping $s_3$ fixed. In doing so we recall that the amplitudes behave as $A\sim \frac{1}{s_i}\log^p s_i$ for $s_i\to\infty$. 
Therefore, in this limit, $\p_{s_i}A$ is suppressed by one additional power. If we furthermore use 
$\p_{s_1}f(s_1)=4\, \bar c/(3 s_1)$ as $s_1\to\infty$, which is 
valid in both phases, we obtain from \eqref{diffeoWI1} the relation
\be
2\,A_1^{\{113\}}-A_4^{\{113\}}=\frac{16\,\bar c}{3\,s_1}
\ee 
We now take the same limit in eq. \eqref{WeylWI1}. This yields
\be
2\,A_1^{\{113\}}-A_4^{\{113\}}=\frac{8\,c}{3\,s_1}
\ee
Comparison gives 
\be
c=2\,\bar c\neq 0
\ee

The normalization of the type A anomaly, \eqref{WeylWI2} cannot be reduced to the two-point function. 
Any regularization respecting diffeomorphism invariance will produce the dimension $-2$ amplitudes corresponding 
to the three energy-momentum correlators  which appear in \eqref{WeylWI2}  and the value 
of $a$ can be simply read off.  In dimensional regularization $a$ is determined by the  $0/0$ contribution 
of a dimension zero amplitude. This amplitudes vanishes in $d=4$ due to the Schouten identity \cite{DS}. 

We will not discuss the dilaton effective action for this case but refer instead to the literature, e.g. \cite{ST}. 

\section{Conclusions}
Our main result is a uniform description of type A and B trace anomalies in $d=4$. As we show the information 
about the anomaly is carried by a Ward identity of the general form
\be\label{eq10000}
s_1\,E_1(s_1,s_2,s_3)+s_2\,E_2(s_2,s_3,s_1)+s_3\,E_3(s_3,s_1,s_2)=ct
\ee
where $s_i\equiv p_{i}^{2}$ are the kinematical invariants ($p_i$ are the three external momenta),
$E_i$ are dimension $-2$  amplitudes, selected depending of the anomaly type and $ct$ is a constant which characterizes 
the strength of the anomaly being respectively related to $a$ or $c$.
The basic Ward identity \eqref{eq10000} can be translated into two equivalent,  universal characterizations of the anomaly:
\be\label{eq10001}
E_i\xrightarrow{s_i\to\infty}{}\frac{ct}{s_i}+{\cal O}\left(\frac{s_j,s_k}{s_i^2}[\log s_i]^p\right)
\ee
and
\be\label{eq10002}
-\frac{1}{\pi}\int d s_i \,{\rm Im}_i E_i(s_i,s_j,s_k)=ct
\ee
where the imaginary part is obtained from the discontinuity with respect to the $s_i$ invariant while the other two invariants $s_j,s_k$ are kept fixed.

After the amplitudes entering the anomaly equation are identified any single one of the conditions \eqref{eq10001}, \eqref{eq10002} implies all the others and also the validity of the basic equation 
\eqref{eq10000} with the same normalization. This depends crucially on the invariant amplitudes having dimension $-2$ and 
obeying the standard analyticity of QFT.
In particular the high invariant behaviour for one of the amplitudes can be related to the two-point
function for type B and to the structure of an invariant amplitude in the three-point function in dimensional regularization which vanishes in $d=4$ for type A.
Once the basic equation \eqref{eq10000} is established trace anomaly matching is immediate: in the deep Euclidean 
limit the invariant amplitudes of the unbroken and broken phases match and since the anomaly is a constant this 
forces $a$ and $c$ to be the same in the two phases.
The basic consequence is then that the anomaly is invariant along the ``flow", i.e. $a,c$ are independent of the 
breaking scale $v$ for the whole range $v=0$, corresponding to the UV unbroken phase, to $v=\infty$, the deep IR of the broken phase. 
This is happening while the individual invariant amplitudes have
a nontrivial dynamical  dependence on the breaking scale $v$ along the ``flow".

Interestingly the same type of equation \eqref{eq10000} is obeyed by chiral anomalies in $d=4$. 
This type of equation generalizes the anomaly information related to ``Dolgov-Zakharov" poles \cite{DZ,BFSY,CG}.
If $s_j,s_k=0$ in \eqref{eq10001}, the sum rule is necessarily saturated by a  $ct \,\delta(s_i)$ singularity signaling a ``pole".
Since, however, the configuration chosen is singular and the amplitudes $E_i$ having branch points at $s_j,s_k=0$, the limit 
to the special configuration should be taken carefully along special lines.
Moreover the ``poles" are effectively representing
a collapsed branch cut or a collision of  two logarithmic branch points in the limit.
The relation between the dimension $-2$ invariant amplitudes appearing in the different anomalies is puzzling. 
In particular the chiral anomaly amplitudes have opposite P and T parities compared with the trace anomaly 
ones and related to that they appear in a phase in the Euclidean configuration space. 
Moreover when the conformal group is extended  to the superconformal one \cite{BPT} they 
appear in the same supermultiplet. Understanding the similarities/differences  of these structures as 
reflected in the equations \eqref{eq10000} obeyed by all of them is an interesting question.

The three-point correlator of energy momentum tensors  can be used to constrain the possible values of the anomalies in unitary CFT  as discussed in \cite{HM}. 
In certain kinematical configurations the correlator reduces to a diagonal matrix element of one energy momentum tensor between two states obtained
by acting on the vacuum with the other two, respectively. It would be interesting to understand if this interpretation carries over for the dimension $-2$ 
amplitudes constraining again their structure.

\section*{Acknowledgement}

We would like to thank Daniele Dorigoni, Lorenzo Casarin and Zohar Komargodski for helpful discussions.
This work was supported in part by an Israel Science Foundation (ISF) center for excellence grant (grant number 2289/18)
\begin{appendices}

\section{Explicit Realizations: Unbroken Phase}

The general discussion presented in Sections 2 and 4 of 
the Ward identities and how the anomaly is captured by the properties of 
dimension $-2$ amplitudes, can be explicitly verified with the 
simplest CFT, namely a free massless scalar field. Here the correlators are one loop Feynman diagrams.  
As the analytic structure of two- and three-point functions is completely fixed by conformal symmetry, 
the results derived for this simple model are universal, the only free parameter being the 
normalization, i.e. the actual strength of the anomaly. 
Furthermore, while the  three-point function of $T_{\mu\nu}$ computed via a one-loop Feynman diagram 
is not identical to $\Gamma^{(3)}_{\mu\nu,\rho\s,\a\b}$, 
the dimensional $-2$ amplitudes can be unambiguously obtained as they are not contaminated by semilocal terms.  

In this appendix we discuss the unbroken phase while in Appendix B we discuss a simple explicit calculable model 
of spontaneous breaking for which the results for  the broken phase can be checked. 

For the conformally coupled scalar with action 
\be
S=\frac{1}{2}\int d^d x\,\sqrt{g}\Big(\nabla^\mu\phi\,\nabla_\mu\phi+\xi\,R\,\phi^2\big)\qquad\qquad
\xi=\frac{d-2}{4(d-1)}
\ee
the on-shell traceless and conserved energy-momentum tensor is 
\be\label{defT}
T_{\mu\nu}(\phi)=\frac{2}{\sqrt{g}}\frac{\delta S}{\delta g^{\mu\nu}}\Big|_{g_{\mu\nu}=\eta_{\mu\nu}}
=\p_\mu\phi\,\p_\nu\phi-\frac{1}{2}\eta_{\mu\nu}\,\p^\rho\phi\,\p_\rho\phi+\xi\big(\eta_{\mu\nu}\square-\p_\mu\p_\nu\big)\phi^2
\ee 
We are interested in $d=4$.  

In the next section, when we discuss the spontaneously broken phase of this simple model, 
we need to consider a massive free scalar. In the Lagrangian the mass term is $-\frac{1}{2}M^2\,\phi^2$
which contributes to the energy-momentum tensor as 
\be
\Delta T_{\mu\nu}=\frac{1}{2}\eta_{\mu\nu}M^2\,\phi^2
\ee
which leads to an explicit breaking of Weyl invariance, i.e. on-shell one how has $T^\mu_\mu=M^2\,\phi^2$.  
In this Appendix we will use these general expressions for $M=0$.

We start with the discussion of the correlator
\be
\langle T_{\mu\nu}(-q)\,{\cal O}(k_1)\,{\cal O}(k_2)\rangle \qquad\hbox{with}\qquad {\cal O}=\phi^2
\ee
The only contributing Feynman diagrams are logarithmically divergent triangle graphs. 
There are two graphs with equal contributions to the amplitudes. 
From our discussion in Section 2 it follows that the divergent part only contributes to the 
amplitude $A$ in the decomposition \eqref{eq9}. This can be easily isolated and the 
finite amplitudes can be recognized by their tensor structures. We assumed that the anomaly appears in 
Weyl invariance; therefore a convenient regularization is dimensional regularization which  
respects diffeomorphism invariance.
The finite,  dimension $-2$ amplitudes are unambiguous, 
not being affected by the contributions of  semi-local terms of the type discussed in Section  3.2.

There are different ways to obtain the finite amplitudes. It turns out that in order to explicitly check the 
features of the invariant amplitudes that we have discussed in Section 2, the most convenient way is the 
Passarino-Veltman \cite{PV} decomposition, which amounts to expressing all Feynman integrals 
with non-trivial tensor structure in terms of basic scalar integrals. This is most easily demonstrated on a 
simple example. Consider the one-loop integral\footnote{The measure for the loop integration has been chosen to 
avoid factors of $4\,\pi$ which can easily be inserted, if needed.},\footnote{With the discussion of Appendix B in mind, 
we treat the massive scalar field.}
\be
B_\mu(p)=\int \frac{d^d l}{\pi^{d/2}} \frac{l_\mu}{(l^2-M^2)((l+p)^2-M^2)}=p_\mu B_1(p)
\ee
where we have used that the index $\mu$ can only be carried by the external momentum $p_\mu$. 
The following simple manipulation 
\be\label{B0}
\ba
p^\mu B_\mu(p)&=\int \frac{d^d l}{\pi^{d/2}}\frac{p\cdot l}{(l^2-M^2)((l+p)^2-M^2)}
=\frac{1}{2}\int \frac{d^d l}{\pi^{d/2}}\frac{((l+p)^2-M^2)-(l^2-M^2)-p^2}{(l^2-M^2)((l+p)^2-M^2)}\\
&=-\frac{1}{2}p^2\,B_0(p)
\ea
\ee
leads to 
\be
B_1(p)=-\frac{1}{2} B_0(p)
\ee
We have used the freedom to shift the loop momentum and we have defined the basic 
scalar two-point one-loop integral 
\be\label{B0}
\ba
B_0(p)&=\int \frac{d^d l}{\pi^{d/2}}\frac{1}{(l^2-M^2)((l+p)^2-M^2)}\\
&=-\frac{2}{d-4}+B_0^f+{\rm const.}+{\cal O}(d-4)
\ea
\ee
where 
\be
B_0^f(p)=-\int_0^1 dx\log \frac {\big(x(1-x)p^2-M^2\big)}{\mu^2}
\ee
and $\mu$ is the arbitrary renormalization scale.
Similarly, one can decompose 
\be
C_{\mu}(k_1,k_2)=\int \frac{d^d l}{\pi^{d/2}}\frac{l_\mu}{((l^2-M^2)((l+k_1)^2-M^2)((l-k_2)^2-M^2)}
\ee
and 
\be
C_{\mu\nu}(k_1,k_2)=\int \frac{d^d l}{\pi^{d/2}}\frac{l_\mu\,l_\nu}{((l^2-M^2)((l+k_1)^2-M^2)((l-k_2)^2-M^2)}
\ee
and express them in terms of $B_0$ and $C_0$ where 
\be\label{C0}
\ba
C_0(k_1,k_2)
=\int \frac{d^d l}{\pi^{d/2}} \frac{1}{\big(l^2-M^2\big)\big((k_1+l)^2-M^2\big)\big((k_2-l)^2-M^2\big)}
\ea
\ee
is the scalar triangle. The tensor indices are now carried by $k_{1\mu}, k_{2\nu}$ and 
$\eta_{\mu\nu}$. 

The final result for the invariant amplitudes $B,C,D$ in $d=4$, which one obtains using the PV decomposition 
of the three-point function $\langle T_{\mu\nu}\,\phi^2\,\phi^2\rangle$ is
\be\label{BCDPV}
\ba
\lambda^4\,B&=2\,r^2\,\lambda^2+\frac{1}{12}\big[q^2\,r^2+2 (\qr)^2\big]\big[3\,q^4-4\,(\qr)^2-2\,q^2\,r^2+3\,r^4\big]C_0
+4\,r^2\,\lambda^2\,M^2\,C_0\\
\noalign{\vskip.2cm}
& -\frac{1}{2}\Big[\big(2(\qr)^3+(q^2-r^2)(2(\qr)^2+q^2 r^2)+\qr(q^2 r^2-3\, r^4)\big)B_0^f(k_1^2)
+(k_1\leftrightarrow k_2)\Big]\\
\noalign{\vskip.2cm}
&-(q^2-r^2)\big[2(\qr)^2+q^2\,r^2\big]B_0^f(q^2)\\
\noalign{\vskip.3cm}
\lambda^4\,C&=-2\,\qr\,\lambda^2-\frac{1}{4}q^2\,\qr\big[3(q^2-r^2)^2-4\,\lambda^2\big]C_0-3\,q^2\,\qr\,(q^2-r^2)B_0^f(q^2)-4\,\qr\,\lambda^2\,M^2\,C_0\\
\noalign{\vskip.2cm}
&+\frac{1}{2}\big[\big((\qr)^2(3\,q^2-r^2)+3\,\qr\,q^2(q^2-r^2)-2\,q^2\,r^4\big)B_0^f(k_1^2)-(k_1\leftrightarrow k_2)\big]\\
\noalign{\vskip.3cm}
\lambda^4\,D&=2\,q^2\,\lambda^2+\frac{1}{4}q^4\big[3(q^2-r^2)^2-4\,\lambda^2\big]C_0+3\,q^4(q^2-r^2)B_0^f(q^2)+4\,q^2\,\lambda^2\,M^2\,C_0\\
\noalign{\vskip.2cm}
&+\frac{1}{2}\big[\big(3\,q^4(r^2-\qr-q^2)-2\,(\qr)^3+5\,\qr\,q^2\,r^2\big)B_0^f(k_1^2)+(k_1\leftrightarrow k_2)\big]
\ea
\ee
where 
\be
\lambda^2=q^4+k_1^4+k_2^4-2\,q^2\,k_1^2-2\,q^2\,k_2^2-2\,k_1^2\,k_2^2
\ee
is the triangle function. 

The expressions for the dimension $-2$ amplitudes above are independent of the renormalization scale $\mu$.
They can be rewritten in terms of differences of logarithms.
Given these explicit expressions for the invariant amplitudes, it is now straightforward 
to check that the Ward identities \eqref{eq17} are satisfied. In the normalization chosen here
\eqref{eq17a} is satisfied with $c=2$ and $\Gamma^{(2)}(k^2)=-2\,\log k^2/\mu^2$, consistent with the general discussion of 
the $\Delta=2$ model in Section 2.

We remark that for the massive scalar the r.h.s. of \eqref{eq17b}  
evaluates to $4+8\,M^2\,C_0$. The additional term reflects the explicit violation of traceless of 
the energy-momentum tensor by the mass term. We will come back to this in Appendix B. 

With the help of \eqref{BCDPV} we can also 
check the asymptotic behaviour of the amplitudes $E_i$. Both in the massless and 
massive cases one finds, as expected, 
\be
E_i\xrightarrow[s_i\to\infty]{}\frac{4}{s_i}
\ee 

Given the expressions for $B,C,D$ we can also check the sum rules. Using the Cutkosky  rule 
one derives e.g., valid for $k_1^2,k_2^2<0$,
\be\label{eq5001}
\ba
&{\rm Im}_{q^2}E_1={\rm Im}_{q^2}(-3\, B-D)\\
&=\pi\Bigg\{-\frac{3(q^2-r^2)(q^4+2\,(\qr)^2+q^2\,r^2)}{\lambda^4}\sqrt{1-\frac{4\,M^2}{q^2}}\\
&+\left(8\,(q^2+3\,r^2)\lambda^2\,M^2+\frac{1}{2}\big(q^4+2\,(\qr)^2+q^2\,r^2\big)
\big(3\,q^4-4\,(\qr)^2-2\,q^2\,r^2+3\,r^4\big)\right)\\
&\qquad\times\frac{1}{\lambda^5}\tanh^{-1}{\left(\frac{\sqrt{1-\tfrac{4 M^2}{q^2}}}{q^2-k_1^2-k_2^2}\,\lambda\right)}\Bigg\}
\theta(q^2-4\,M^2)
\ea
\ee
From this one computes
\be\label{eq5002}
-\frac{1}{\pi}\int_{4\,M^2}^\infty dq^2\,{\rm Im}_{q^2}E_1(q^2,k_1^2,k_2^2,M^2)=4
\ee
For later use we have again presented the results for a massive scalar but, of course the result being 
independent of $M$, it is also valid for the discontinuity evaluated at $M=0$. 

The computation of the correlation function of three energy-momentum tensors 
\be
\langle T_{\mu\nu}(k_1)\,T_{\rho\s}(k_2)\,T_{\a\b}(k_3)\rangle\qquad\qquad\hbox{with}\qquad k_1+k_2+k_3=0
\ee
takes more effort. Rather than doing a Passarino-Veltman decomposition, 
we have derived for the 27 dimension $-2$ amplitudes expressions involving integration of the 
two Feynman parameters (we work again in dimensional regularization). They all have the form 
\be\label{Feynman_integral}
\int_0^1 dx\int_0^{1-x}\!\!\!\!\!dy\,\,\frac{P(x,y)}{x\,y\,k_3^2+x(1-x-y)k_1^2+y(1-x-y)k_2^2-M^2}
\ee
where $P(x,y)$ are polynomials in the Feynman parameters. For the unbroken phase which 
we discuss here, $M^2=0$. 

The calculation is straightforward, however the detailed results are too long to present here. 
But they were used to check all the Ward identities 
which we have written in Section 4, in particular that the combination of amplitudes 
in the Weyl Ward identities are constants, independent of the kinematical invariants. 
Also the diffeomorphism Ward identity \eqref{diffeoWI1} has been verified in this way. More precisely, the 
Ward identities are satisfied in this simple model  
for $(4\pi)^2\,c=\frac{1}{120}$ and $(4\pi)^2\,a=\frac{1}{360}$, which are known values for the free scalar; see e.g. \cite{Duff}. 

\section{Explicit Realizations: Broken Phase}
In this Appendix we check the general setup for the anomaly structure in the broken phase within a simple model proposed in \cite{KS}.
Consider two massless scalar fields $\phi$ and $\varphi$ interacting through a marginal perturbation:
\be\label{eq5000}
\textit{L}=\frac{1}{2}\partial_{\mu}\phi\, \partial^{\mu}\phi +\frac{1}{2}\partial_{\mu}\varphi\, \partial^{\mu}\varphi-g\,\phi^2\,\varphi^2
\ee
The fields are coupled conformally to a background metric.
Both in the unbroken and broken phases we will take a limit where $g$ goes to $0$ and 
therefore the beta function(s) vanish, thus not disturbing conformality.
Therefore the unbroken phase is made simply from two decoupled massless scalar fields.
Consider now the breaking: we give the field $\varphi$ a vacuum expectation value $v$, i.e.
\be\label{eq11p}
\langle\varphi\rangle=v
\ee
In order to calculate in the broken vacuum, we can alternatively shift in the 
Lagrangian the field $\varphi$
\be\label{eq12p}
\varphi=v+\tilde\varphi
\ee
and calculate with the usual Feynman rules  for the field $\tilde \varphi$ which has zero
$vev$. The dimensionless dilaton  $\Sigma$, which transforms linearly under Weyl transformations, is
\be\label{eq700}
\Sigma=\log\left(1+\frac{\tilde\varphi}{v}\right)\simeq \frac{\tilde\varphi}{v}+{\cal O}(\varphi^2)
\ee
starting linearly in $\tilde\varphi$.
Since the original energy-momentum tensor is
\be\label{eq13p}
\ba
T_{\mu\nu}(\phi,\varphi)=T_{\mu\nu}(\phi)+T_{\mu\nu}(\varphi)+\frac{1}{2}\eta_{\mu\nu}\,g\,\phi^2\,\varphi^2
\ea
\ee
the shift produces a linear coupling of the dilaton in the energy-momentum tensor, 
\be\label{DeltaT}
\frac{1}{3}v^2\big(\eta_{\mu\nu}\square-\p_\mu\p_\nu\big)\Sigma
\ee 
which leads to
\be\label{eq14p}
\langle0|T_{\mu\nu}|\Sigma(p)\rangle=\frac{1}{3}v^2\,p_\mu\,p_\nu
\ee
Covariantly the above coupling is translated into a $v^2\, \Sigma\, R$ term in the effective Lagrangian, 
where $R$ is the curvature scalar.
Also a mass term for the $\phi$ field with $M^2=2\,g\,v^2$  and a cubic coupling 
$-2 M^2\, \Sigma\, \phi^2$ are produced.

We will take the limit 
\be\label{eq15p}
g\to 0\,,\qquad v\to\infty\,,\qquad M^2=2\, g\, v^2=\hbox{fixed}
\ee

The dimension $2$ operator will be
\be\label{eq10p}
{\cal O}(x)=\phi^2(x)
\ee

The broken phase is defined by the Feynman diagrams which survive this limit.
All the correlators of the $\phi^2$  operators and energy-momentum tensors coupled directly or through the  dilaton  
have a scale $M$. This is the consequence of the dilaton having the propagator proportional to $\frac{1}{v^2}$ 
which cancels $v^2$ in the dilaton coupling to the scalar curvature.
 
We recapitulate the content of the broken phase: 
\begin{itemize}
\item[a)] 
A massive scalar field $\phi$ with mass $M$  with standard massive propagator and  an energy-momentum tensor 
containing the conformal improvement and the mass term.
\item[b)]
A massless dilaton field $\Sigma$ with propagator normalized to $\frac{1}{v^2}$.
\item[c)]
The dilaton is coupled to the massive field through a $M^2\, \Sigma\,\phi^2$ coupling and to the energy-momentum tensor 
by a $v^2 p_{\mu} p_{\nu}$ coupling. 
All the diagrams involving correlators of the massive energy-momentum tensor and operators made of the massive field have the scale $M$ 
and are well defined in the limit.
\item[d)]
The dilaton has a Weyl invariant kinetic term inherited from the $\varphi$ field. Its energy-momentum tensor is decoupled from 
the rest of the system and has the $v$ independent trace anomalies expected for  a free massless field. The kinetic term contains 
dilaton self-interactions with the scale $v$ which goes to $\infty$
in the limit considered, but being decoupled we will ignore this sector. 
\end{itemize}
 
We start with the discussion of the $\Delta=2$ model in this particular broken phase.
The two-point function is simply 
the mass term corrrelator for a massive field, i.e. it is logarithmically divergent.
After renormalization it is given by
\be\label{eqref16p}
\Gamma^{(2)}(p^2)=\Gamma^{(2)}(\mu^2)+\frac{1}{(4\,\pi)^2}(p^2-\mu^2)\int_{4M^2}^\infty
dx\frac{\sqrt{1-\frac{4M^2}{x}}}{(x-p^2)(x-\mu^2)}
\ee
Its exact form will not play a role in our calculation.
As discussed in Section 2, after using diffeomorphism invariance the logarithmically divergent contributions of the two and three-point 
functions drop out and we are left with Ward identity \eqref{eq1000} which involves only the dimension $-2$ amplitudes of the   
$\langle T_{\mu\nu}\, {\cal O}\, {\cal O}\rangle$ 
correlator in the broken phase. In the limit \eqref{eq15p} there are two diagrams 
which survive (see Figure 2)
\begin{figure}[htb]
\centering
\begin{tikzpicture}[scale=1.5]
\draw[style=thick](0,0)--(1.2,1) node [shift={(0.4,.0)}] {$\cal{O}$};
\draw[style=thick](0,0)--(1.2,-1) node [shift={(0.4,.0)}] {$\cal{O}$};
\draw[style=thick](1.2,1)--(1.2,-1) node [shift={(-2.0,1.9)}] {$-2gv$} ;
\draw node [shift={(0.0,0,0)}] {$\bullet$}; 
\draw node [shift={(-1.4,0.0)}] {$\bullet$};
\draw node [shift={(-1.4,0.3)}] {$v$}; 
\draw[style=dashed](0,0)--(-1.0,0) node [shift={(-0.6,.0)}] {$T_{\mu\nu}$};
\end{tikzpicture}
\begin{tikzpicture}[scale=1.5]
\draw[style=thick](0,0)--(1.2,1) node [shift={(0.4,.0)}] {$\cal{O}$};
\draw[style=thick](0,0)--(1.2,-1) node [shift={(0.4,.0)}] {$\cal{O}$};
\draw[style=thick](1.2,1)--(1.2,-1);
\draw node [shift={(0.0,0,0)}] {$\bullet$}; 
\draw node [shift={(-0.5,.0)}] {$T_{\mu\nu}$};
\draw node [shift={(-1.7,.0)}] {$+$};
\end{tikzpicture}
\caption{}
\end{figure}
corresponding to the coupling of the energy-momentum tensor through the dilaton and directly.
Taking the trace of the energy-momentum tensor gives the combination entering \eqref{eq1000} whose right hand side is 
the anomaly, a constant independent on the kinematical invariants and the scale $M$. We remark that the dilaton contribution 
to the trace is, with opposite sign, equal to the contribution of the correlator of $M^2\,\phi^2$ with two ${\cal O}$ operators. 
Therefore an alternative interpretation of the anomaly equation in this very special broken phase is that it represents 
an anomaly in the Ward identity satisfied by the trace of a free massive  scalar 
\be\label{eq17p}
T^\mu_\mu-M^2{\cal O}\simeq 0
\ee
which is valid for the  free massive scalar with energy-momentum tensor
\be\label{eq18p}
T_{\mu\nu}=\p_\mu\phi\,\p_\nu\phi-\frac{1}{2}\eta_{\mu\nu}\,\p^\a\phi\,\p_\a\phi
+\frac{1}{6}\big(\eta_{\mu\nu}\square-\p_\mu\p_\nu\big)\phi^2+\frac{1}{2}\eta_{\mu\nu}\,M^2\,\phi^2
\ee
evaluated in a correlator with two ${\cal O}$ operators.

For our very simple model the limit \eqref{eq15p} selected the diagrams which participate in the relevant Ward identities: 
the two diagrams of Figure 2 and the two-point function \eqref{eqref16p}. Then using the regularization which respects diffeomorphism invariance one arrives at \eqref{eq17b} 
which defines the anomaly in the spontaneously 
broken phase in terms of the dimension $-2$ amplitudes  contained in the two diagrams  of Figure 2. 
Without such a complete analysis even the 
meaning of the anomaly in the spontaneously broken phase is not clear.
In particular  we intend to analyze in the future  if the 
interesting models with potential control of the broken phase proposed in refs. \cite{Pomoni1,Pomoni2} can be brought to this level of meaningful analysis.              
We continue now the analysis of our simple model.

We can now check the Weyl Ward identity \eqref{eq17b} in the broken phase for this  model. This requires the knowledge 
of the dimension $-2$ amplitudes contributed by the two diagrams. 

The  first diagram, using the linear dilaton coupling \eqref{DeltaT}, gives a contribution $\Delta B$ to the amplitude $B$
\be\label{DeltaB}
\Delta B=-\frac{8}{3}\frac{M^2}{q^2}\,C_0
\ee 
with $C_0$ as in \eqref{C0}. $C_0$ and the amplitudes $B,C,D$ corresponding to the second diagram  are now those for the massive case. 
The amplitudes $C$ and $D$ are not modified. 
We gave in \eqref{eq5001} the expression for the contribution to the discontinuity in $q^2$ of the $E_1$ amplitude from the second diagram. 
The contribution to the sum rule from the first diagram, i.e. the contribution from  $\Delta B$, is zero. The reason is simply that, at 
high $q^2$, $\Delta B$ behaves as $\frac{(\log q^2)^2} {(q^2)^2}$ with the power $\frac{1}{q^2}$, whose coefficient is the sum rule contribution, missing. 
Therefore at finite $M$ the anomaly is controlled by the second diagram in Figure 2 which, as shown in the 
previous Appendix, saturates the sum-rule. This verifies anomaly matching explicitly.  

We discuss now the anomaly in the IR limit of the broken phase, i.e. when $M$ goes to $\infty$.
Since one takes the limit of  $M$ first we cannot consider anymore the high momentum behaviour 
of the amplitudes or the sum rules derived from them.
We should use instead directly \eqref{eq400}.  As the anomaly is independent of $M$ 
we expect the matching to work also at $M=\infty$. 
The second diagram vanishes in this limit (we remind that we are discussing all the time the dimension $-2$ amplitudes).
We have to evaluate the first, i.e. the dilaton diagram. It has a finite limit giving
\be\label{eq2005}
E_1=\frac {4}{q^2}
\ee
while $E_2=E_3=0$. Therefore the anomaly equation \eqref{eq400} is satisfied with $c^B=c=2$. 
Since the matching happened due to the specific value of the dilaton coupling to two ${\cal O}$ operators 
we see explicitly the connection between the anomaly matching and the constraints on dilaton couplings.

We comment on two additional features of this calculation:

a) The same limit appears in the calculation of the anomaly of the massless scalar field when 
Pauli-Villars regularization is used. Then the trace of the energy-momentum tensor in a correlator with two ${\cal O}$ operators 
is given by the explicit violation introduced by the Pauli-Villars regulator.
Therefore the limit with opposite sign represents the anomaly.

b) As discussed above, in this simple model the anomaly in the broken phase for finite $M$ is related to the anomaly of a 
massive scalar. One can relate therefore the spontaneous breaking in the conformal theory to a ``massive flow" specifically 
of the $\phi$ scalar which starts massless in the UV and in the IR has an infinite mass. As we described above for finite $M$ 
one had the anomaly in the correlator of the energy-momentum tensor of the massive scalar with two ${\cal O}$ operators. 
At $M=\infty$ this correlator vanishes  and therefore the anomaly in the IR is zero. Hence, from the-point of 
view of the massive flow one has different anomalies in the UV and IR. In the broken CFT description one has 
anomaly matching and a physical dilaton degree of freedom in the IR.  As a consequence the first,
i.e. the dilaton diagram, makes up the difference as calculated above.  Therefore generally for a massive flow 
it is natural to describe it  in terms of a dilaton source (not a physical state) 
which contains  the structure  
of the nonvanishing difference between the UV and IR anomalies on the 
massive flow \cite{KS,Komargodski,Luty}.\footnote{For a bootstrap approach see \cite{Penedones}.}

One can use the same model to verify the general results presented in Section 4 for the 
correlator of three energy-momentum tensors in the broken phase. This is much more involved and 
we discuss here only the analysis of the $M\to\infty$ limit.
We start with the discussion of the general set up. In the broken phase we should consider for the three-point 
function all the contributions where the energy-momentum tensor 
couples directly to the massive loop or through up to three dilatons.
Since we are interested in the anomalous part of the effective action, in principle we should isolate
the contributions of the dimension $-2$ amplitudes entering the anomalous equations, i.e.
\eqref{WeylWI1} and \eqref{WeylWI2}. 
In the $M\to\infty$ limit the contribution of the diagram with direct couplings of all 
three energy-momentum tensors vanishes for dimension $-2$ amplitudes.

For the diagrams  where the energy-momentum tensor couples through at least one dilaton we use a short cut.
Locality of the anomaly implies that in order that the diagram contributes to the anomalous Ward identity its expression 
should have exactly one dilaton propagator. Therefore in the $M\to\infty$ limit, after factoring out the propagator,
the rest of the diagram should give a dimensionless coefficient multiplying four momenta. 
The interpretation of the coefficient is that of the  normalization of a two-dilaton -- one-metric perturbation   
or three-dilaton terms in the anomalous  Wess-Zumino action.
In order to generate these terms in the limit one has to expand the respective triangle Feynman diagram 
in the  momenta carried by the additional dilatons. The momenta cancel the additional propagators and 
the $M\to\infty$ limit is finite.

In addition the diagrams with no external momenta have also positive powers of $M$ in the expansion. These terms  
are non-anomalous  since after taking the trace of the energy-momentum tensor they have dilaton propagators 
and of course the anomaly cannot have such an analytic structure.
These effective non-anomalous tree diagrams which involve dilaton propagators arise from the 
non-anomalous kinetic term of the dilaton.  
Therefore the contributions with positive powers of $M$ 
in the limit $M\to\infty$ can be included as a ``renormalization" of the kinetic term. 

To summarize, the anomalous contribution of the diagrams with couplings through the dilaton is 
the finite contribution (through expansion in momenta) in the $M\to\infty$ limit.   
These anomalous contributions for $2,3,4$ dilatons were calculated in \cite{KS}.
Here we complete the calculation for the single dilaton which gives the linear 
coupling of the dilaton to the anomaly curvature polynomials. By our discussion above, 
this is captured by the left diagram in Figure 2 where the two operators ${\cal O}$ are replaced 
by energy-momentum tensors. This amounts to computing the correlator
\be\label{OTT}
\langle M^2\,\phi^2(q)\,T_{\mu\nu}(k_1)\,T_{\rho\s}(k_2)\rangle
\ee
More specifically we computed the unambiguous finite part of this dimension $+2$ correlator, where all 
four tensor indices are carried by the two momenta $k_1$ and $k_2$. The invariant amplitudes have the 
general form  \eqref{Feynman_integral}, but in the limit $M^2\to\infty$ we can replace the denominator 
by $-M^2$. In this case the integral over the two Feynman parameters becomes trivial. 
This (finite) part of \eqref{OTT} should be compared with the ${\cal O}(h^2)$ expansion of the anomaly 
\eqref{WeylAnomaly}. In fact, we keep from it only the piece where all four tensor indices are 
carried by the momenta of the two gravitons $h^{\mu\nu}$. This leads to an over-determined system of 
linear equations which is solved by  $(4\pi)^2(a,c,b)=\frac{1}{360}(1,3,2)$ as expected \cite{Duff}.  

In addition to the contribution to the anomaly we have discussed so far, there is also 
the dilaton loop which contributes with equal coefficients as $\phi$ such that the 
total anomaly is that of two free scalars.  This dilaton contribution is generic and 
not special to this simple model. 

\end{appendices}

\end{document}